\documentclass[aos,MSNbibl,seceqn,dvips]{arximspdf}
\usepackage{graphicx}
\doi{10.1214/14-AOS1278} % kopijuoti is laisko
\volume{43}
\issue{1}
\pubyear{2015}
\firstpage{238}
\lastpage{275}
\docsubty{FLA}

\makeatletter
\newcommand{\ds}{\displaystyle}
\newcommand{\lleft}{\left}
\newcommand{\rrvert}{\vert}
\newcommand{\rright}{\right}
\newcommand{\llvert}{\vert}
\newcommand{\eqref}[1]{(\ref{#1})}
\newtheorem{theorem}{Theorem}
\newtheorem{lemma}{Lemma}
\newproclaim{remarks}{Remarks}
\newproclaim{remark}{Remark}
\newtheorem{prop}{Proposition}
\newtheorem{cor}{Corollary}
\newproclaim{assumptions}{Assumption}
\makeatother

\begin{document}
\begin{frontmatter}

\title{On the efficiency of pseudo-marginal random walk Metropolis algorithms}
\runtitle{Efficiency of pseudo-marginal RWM algorithms}

\begin{aug}
\author[A]{\fnms{Chris}~\snm{Sherlock}\corref{}\ead[label=e1]{c.sherlock@lancaster.ac.uk}},
\author[B]{\fnms{Alexandre H.}~\snm{Thiery}\ead[label=e2]{a.h.thiery@nus.edu.sg}},
\author[C]{\fnms{Gareth~O.}~\snm{Roberts}\ead[label=e3]{Gareth.O.Roberts@warwick.ac.uk}} 
\and
\author[D]{\fnms{Jeffrey S.}~\snm{Rosenthal}\ead[label=e4]{jeff@math.toronto.edu}}
\runauthor{Sherlock, Thiery, Roberts and Rosenthal}
\affiliation{Lancaster University, National University of Singapore,\\
University of Warwick and University of Toronto}
\address[A]{C. Sherlock\\
Department of Mathematics\\
\quad and Statistics\\
Lancaster University\\
Lancaster LA1 4YF\\
United Kingdom\\
\printead{e1}}
\address[B]{A. H. Thiery\\
Department of Statistics\\
\quad and Applied Probability\\
Faculty of Science\\
National University\\
\quad of Singapore (NUS)\\
Singapore 117546\\
\printead{e2}}
\address[C]{G. O. Roberts\\
Department of Statistics\\
University of Warwick\\
Coventry CV4 7AL\\
United Kingdom \\
\printead{e3}}
\address[D]{J. S. Rosenthal\\
Department of Statistics\\
University of Toronto\\
100 St. George Street\\
Toronto, Ontario M5S 3G3\\
Canada\\
\printead{e4}}
\end{aug}

% HISTORY:
\received{\smonth{9} \syear{2013}}
\revised{\smonth{10} \syear{2014}}

% ABSTRACT
%
\begin{abstract}
We examine the behaviour of the pseudo-marginal
random walk Me\-tropolis algorithm, where evaluations of the target
density for the accept/reject probability are estimated rather than
computed precisely.
Under relatively general conditions on the target distribution,
we obtain limiting formulae for the acceptance rate and for the
expected squared jump distance, as the dimension of the
target approaches infinity, under the assumption that the noise in the
estimate of the log-target is additive and is independent of the
position.
For targets with independent and identically distributed components,
we also obtain a limiting diffusion for the first component.

We then consider the overall efficiency of the algorithm, in terms of
both speed of mixing and computational time. Assuming the additive
noise is Gaussian and is inversely proportional to the number of
unbiased estimates that are used, we prove that the algorithm is
optimally efficient when the variance of the noise is approximately
3.283 and the acceptance rate is approximately 7.001\%. We also
find that the optimal scaling is insensitive to the noise and that the
optimal variance of the noise is insensitive to the scaling. The theory is
illustrated with a simulation study using the particle marginal
random walk Metropolis.
\end{abstract}

% KEYWORDS
\end{frontmatter}

%s1 #&#
\section{Introduction}

Markov chain Monte Carlo (MCMC) algorithms have proved
particularly successful in statistics for investigating posterior
distributions in Bayesian analysis of complex models; see, for
example, \cite{RobSmi93,Tierney1994,MCMChandbook}.
Almost all MCMC methods are based on the Metropolis--Hastings (MH) algorithm
which owes much of its success to its tremendous flexibility. However,
in order to use the classical MH algorithm, it must be
possible to evaluate the target density up to a fixed constant of
proportionality. While this is often possible, it is increasingly
common for
exact pointwise likelihood evaluation to be prohibitively expensive, perhaps
due to the sheer size of the data set being analysed. In these situations,
classical MH is rendered inapplicable.

The \textit{pseudo-marginal Metropolis--Hastings algorithm} (PsMMH)
\cite{beau03,AndrieuRoberts2009} provides a general recipe for
circumventing the need for target density evaluation. Instead it is
required only to be able to unbiasedly \textit{estimate} this density.
%allows such computations when the target
%cannot be evaluated up to a constant of proportionality, provided that
%an unbiased (up to a fixed constant of proportionality) estimator of
%the
%target density is available. It has enabled mankind
%to move to a higher plane \cscomment{Gareth: refs? expand? contract?}
%...
The target densities in the numerator and
denominator of the MH accept/reject ratio are then replaced
by their unbiased estimates.
Remarkably, this yields an algorithm which still has the target as its
invariant
distribution. One possible choice of algorithm, the
\textit{pseudo-marginal random walk Metropolis} (PsMRWM),
is popular
in practice (e.g., \cite{GolightlyWilkinson2011,KnapedeValpine2012})
because it requires
no further information about the target, such as the local gradient or
Hessian, which are
generally more computationally expensive to approximate than the
target itself \cite{Poyiadjisetal2011}.

Broadly speaking, the mixing rate of any PsMMH algorithm
decreases as the dispersion in the estimation of
the target density increases \cite{AndrieuRoberts2009}.
In particular, if the target density happens to be
substantially over-estimated, then the chain will be overly reluctant
to move from that state leading to a long run of successive rejections
(a \textit{sticky patch}).
Now, in PsMMH algorithms, the target estimate is
usually computed using an average of
some number, $m$, of approximations; see Sections~\ref{sect.psrwm.intro}~and~\ref{sect.optimising}.
This leads to a trade off, with increasing $m$ leading to
better mixing of the chain, but also to larger computational expense.
We shall consider the problem of optimising $m$.

It is well known (e.g., \cite{RobertsRosenthal2001,SherlockFearnheadRoberts2010}) that the efficiency of the
random-walk Metropolis (RWM) algorithm varies enormously with the
scale of the proposed jumps. Small proposed jumps lead to high acceptance
rates but little movement
across the state space, whereas large proposed jumps lead to low
acceptance rates and again to inefficient exploration of the
state space. The problem of choosing the optimal scale of the RWM proposal
has been tackled for various shapes of target
(e.g., \cite{RobertsGelmanGilks1997,RobertsRosenthal2001,BedardA2007,BedardRosenthal2008,BeskosRobertsStuart2009,SherlockRoberts2009,Sherlock2013,BreRob00})
and has led to the following rule of thumb: choose the scale so that the
acceptance rate is approximately $0.234$. Although nearly all of the
theoretical results are based upon limiting arguments in high
dimension, the rule of thumb appears to be applicable even in
relatively low dimensions
(e.g., \cite{SherlockFearnheadRoberts2010}).
%, \cscomment{Gareth: more refs, e.g. other Mylene?}).

This article focusses on the efficiency of the PsMRWM as the
dimension of the target density diverges to infinity.
For relatively general forms of the target distribution,
under the assumption of additive independent noise in the log-target,
we obtain (Theorem~\ref{thm.asymp.analysis}) expressions
for the limiting expected squared jump distance (ESJD) and asymptotic
acceptance rate.
ESJD is now well established as a pragmatic and useful measure of
mixing for MCMC algorithms in many contexts (see, e.g., \cite{PasGel10}),
and is particularly relevant when diffusion limits can be established;
see, for example, the discussion in \cite{RobRos12}.
We then prove a diffusion limit for a rescaling of the first
component, in the case of a target with independent and
identically distributed components (Theorem~\ref{thm.diff.lim}),
the efficiency of the algorithm is then given by the
speed of this limiting diffusion, which is equivalent to the limiting ESJD.
We examine the relationship between efficiency,
scaling, and the distributional form of the noise, and
consider the \textit{joint} optimisation of the efficiency of
the PsMRWM algorithm (taking computational time into account)
with respect to~$m$,
and the RWM scale parameter. Exact analytical results are
obtained (Corollary~\ref{cor.max.Gauss}) under an assumption of
Gaussian noise in the estimate of the log-target, with a~variance that
is inversely proportional to $m$. In this case, we prove that the
optimal noise variance is 3.283, and the corresponding optimal asymptotic
acceptance rate is 7.001\%, thus extending the previous 23.4\% result of
\cite{RobertsGelmanGilks1997}.
%Although our theorems assume that $d\to\infty$, we illustrate
%(Section~\ref{sect.low.dimension}) that they remain approximately true
%even in moderate dimension.
Finally, we illustrate the use of these theoretical results
in a simulation study (Section~\ref{sect.sim.study}).

%s1.1 #&#
\subsection{The PsMRWM}
\label{sect.psrwm.intro}

Consider a state space $\mathcal{X}\subseteq\mathbb{R}^d$, and let
$\pi(\cdot)$ be a
distribution on $\mathcal{X}$, whose density (with respect to Lebesgue measure)
will be referred to as $\pi(\mathbf{x})$.
The MH updating scheme provides a very general class
of algorithms for obtaining an approximate dependent sample from a target
distribution, $\pi(\cdot)$, by constructing a Markov chain with $\pi(\cdot)$
as its limiting distribution.
Given the current value $\mathbf{x}$, a new value $\mathbf{x}^*$ is proposed
from a pre-specified Lebesgue density $q (\mathbf{x},\mathbf{x}^*)$ and is then
accepted with probability
$\alpha(\mathbf{x},\mathbf{x}^*) =
1 \wedge
[\pi(\mathbf{x}^*)  q (\mathbf{x}^*, \mathbf{x}
)]/[\pi(\mathbf{x})  q (\mathbf{x},\mathbf{x}^*)]$.
If the proposed value is accepted, then it becomes the next current value;
otherwise the current value is left unchanged.

The PsMMH algorithm
\cite{AndrieuRoberts2009} presumes the computational
infeasibility of evaluating $\pi(\mathbf{x})$ and uses an approximation
$\hat{\pi}_\mathbf{v}(\mathbf{x})$ that depends on some auxiliary
variable, $\mathbf{v}$.
The auxiliary variable is sampled from some distribution
$q_{\mathrm{aux}}(\mathbf{v}|\mathbf{x})$, and the approximation
$\hat{\pi}_\mathbf{v}(\mathbf{x})$
is assumed to satisfy that
$\mathbb{E}_{q_{\mathrm{aux}}}[\hat{\pi}_{\mathbf{V}}(\mathbf{x})]=c\pi
(\mathbf{x})$, for some
constant $c>0$. The value of the constant is irrelevant to all that
follows, and so, without loss of generality, we assume that $c=1$.
%Perhaps surprisingly, the PsMMH algorithm can function even when $
We also assume that $\hat{\pi}_\mathbf{v}>0$.

The PsMMH algorithm creates a Markov chain with a stationary
density (since $c=1$) of
%
%e1.1 #&#
\begin{equation}
\label{eqn.joint.stat.v}
\tilde{\pi}(\mathbf{x},\mathbf{v})= q_{\mathrm{aux}}(\mathbf{x},
\mathbf{v})\hat{\pi}_\mathbf {v}(\mathbf{x}),
\end{equation}
which has $\pi(\mathbf{x})$ as its $\mathbf{x}$ marginal.
When a new value, $\mathbf{X}^*$, is proposed via the MH algorithm, a
new auxiliary
variable, $\mathbf{V}^*$, is proposed from the density
$q_{\mathrm{aux}}(\mathbf{x}^*,\mathbf{v}^*)$. The pair $(\mathbf{x}^*,\mathbf{v}^*)$ are then
jointly accepted or rejected. The acceptance probability for this
MH algorithm on $(\mathbf{x},\mathbf{v})$ is
%$1\wedge R$, where
\[
%R=
%{\tilde{\pi}(\bmx,\bmv)q(\bmx,\bmx^*)q_{\mathrm{aux}}(\bmx^*,\bmv^*)}
%=
1 \wedge \frac{\hat{\pi}_{\mathbf{v}^*}(\mathbf{x}^*) q (\mathbf{x}^*, \mathbf{x})}{
\hat{\pi}_{\mathbf{v}}(\mathbf{x})
q ({\mathbf{x},\mathbf{x}^*})}.
\]
We are thus able to substitute the estimated density for the true
density, and still obtain the desired stationary distribution for
$\mathbf{x}$.
Note that for symmetric proposals, this simplifies to
$1\wedge  [ \hat{\pi}_{\mathbf{v}^*}(\mathbf{x}^*)/\hat{\pi
}_{\mathbf{v}}(\mathbf{x})]$.

%and
%ways of getting the unbiased estimator (importance sampling, particle
%filters,
%poisson estimator a la Beskos/Fernhead/Roberts, ..)]}

Different strategies exist for
producing unbiased estimators, for instance, using
importance sampling or latent variable
representations, as in \cite{MR2523903}, or using particle
filters \cite{DelMoral2004,GordonSalmondSmith1993} as
in \cite{AndrieuDoucetHolenstein2010}. We shall illustrate our theory in
the context of Bayesian analysis of a partially observed Markov jump
process.

%s1.2 #&#
\subsection{Previous related literature}
\label{sect.psrwm.literature}

Pitt et al. \cite{Pittetal2012} and Doucet et al. \cite
{Doucetetal2014} examine the efficiency of pseudo-marginal
algorithms using bounds on the integrated autocorrelation time
($I_{\mathrm{ACT}}$) and under
the assumptions that the chain is stationary
and the distribution of the additive noise in the
log-target is independent of $\mathbf{x}$ (our Assumption~\ref{ass.noise.diff.indep}). Under the
further assumption that this additive noise is Gaussian and the
computing time inversely proportional to its variance (our
Assumption~\ref{ass.standard.regime}), both articles
then seek information on the optimal variance of this additive
noise. Pitt et al. \cite{Pittetal2012} consider the (unrealistic)
case where the Metropolis--Hastings
algorithm is an independence sampler which proposes from the desired
target distribution for~$x$, and obtain an optimal variance of
$0.92^2$. Doucet et al. \cite{Doucetetal2014} consider a general
Metropolis--Hastings algorithm and define a parallel hypothetical
kernel $Q^*$ with the same proposal mechanism as the original kernel,
$Q$, but where the
acceptance rate separates into the product of that of the idealised marginal
algorithm (if the true target were known) and that of an independence
sampler which proposes from the assumed distribution for the
noise. This kernel can never be more efficient than the true kernel.
Upper and lower bounds are obtained for the $I_{\mathrm{ACT}}$ for $Q^*$ in
terms of the
of $I_{\mathrm{ACT}}$ of the exact chain and the $I_{\mathrm{ACT}}$ and a particular
lag-1 autocorrelation of the independence
sampler on the noise. These bounds are examined under the assumption
that the additive noise is Gaussian and the optimal variance for the
noise is estimated to lie between $0.92^2$ and $1.68^2$.

Other theoretical properties of pseudo-marginal algorithms are
considered in~\cite{AndrieuVihola2014}, which gives qualitative
(geometric and polynomial ergodicity) results for the method and some
results concerning the loss in efficiency caused by having to estimate the
target density.

%s1.3 #&#
\subsection{Notation}

In this paper, we follow the standard convention whereby capital
letters denote random variables, and lower case letters denote their
actual values. Bold characters are used to denote vectors or matrices.

%s2 #&#
\section{Studying the pseudo marginal random walk Metropolis in high
dimensions}

%In section [] we state our assumptions on the sequence of target
%distributions $\pi^{(d)}$.

% \subsection{Random Walk Metropolis}
%s2.1 #&#
\subsection{Proposal distribution}

We focus on the case where the proposal, $\mathbf{x}^*$, for an update
to $\mathbf{x}$ is
assumed to arise from a
random walk Metropolis algorithm with an isotropic Gaussian proposal
%
%e2.1 #&#
\begin{equation}
\label{eqn.rwm.prop}
\mathbf{X}^*=\mathbf{x}+\lambda\mathbf{Z} \qquad \mbox{where }  \mathbf{Z}
\stackrel{\mathcal{D}} {\sim }\mathbf{N}(\mathbf{0},\mathbf{I}),
\end{equation}
and $\mathbf{I}$ is the $d\times d$
identity matrix, and $\lambda> 0$ is the scaling parameter for the
proposal. The results presented in this article extend easily to a more general
correlation matrix by simply considering the linear co-ordinate transformation
which maps this correlation matrix to the identity matrix and
examining the target in this transformed space.
In proving the limiting results we consider a sequence of
$d$-dimensional target probabilities $\pi^{(d)}$.
%random variables, $\left\{\bmxc{d}\right\}~(d=1,2,\dots)$ with
%distribution $\pi^{(d)}(\cdot)$.
In dimension $d$ the proposal is $\mathbf{X}^{(d)*} \stackrel{\mathcal
{D}}{\sim}\mathbf{N}(\mathbf{x}^{({d})},\lambda^{(d)2}\mathbf{I}^{(d)})$.

%s2.2 #&#
\subsection{Noise in the estimate of the log-target}

We will work throughout with the log-density of the target,
and it will be convenient to consider the
difference between the estimated log-target
[$\log\hat\pi_V(\mathbf{x})$]
and the true log-target [$\log\pi(\mathbf{x})$]
at both the proposed values ($\mathbf{x}^*,V^*$) and the current
values ($\mathbf{x},V$),
as well as the difference between these two differences,
%
%e2.2 #&#
\begin{equation}
\label{WBdef}
\cases{\ds W:=\log\hat{\pi}_V(
\mathbf{x})-\log\pi(\mathbf{x}),
\vspace*{3pt}\cr
\ds W^*:=\log\hat{\pi}_{V^*} \bigl(\mathbf{x}^* \bigr)-\log\pi \bigl(
\mathbf{x}^* \bigr),
\vspace*{3pt}\cr
\ds B:=W^*-W.}
\end{equation}
Throughout this article we assume the following.

%as1 #&#
\begin{assumptions}
\label{ass.noise.diff.indep}
The Markov chain $(\mathbf{X},W)= \{ (\mathbf{X}_k,W_k)  \}
_{k\geq0}$ is
stationary, and the distribution of the additive noise in the
estimated log-target at the proposal,~$W^*$, is independent
of the proposal itself, $\mathbf{X}^*$.
\end{assumptions}
%
%Unless otherwise stated, we will thus always assume that the different
%Markov chains considered in this article evolve in stationarity.

%re1 #&#
\begin{remark}\label{rem1}
It is unrealistic to believe that the second part of Assumption~\ref{ass.noise.diff.indep} should hold in practice. Pragmatically,
this assumption is necessary in order to make progress with the
theory presented herein; however, in our simulation study
in Section~\ref{sect.sim.study} we provide evidence that, in the scenarios
considered, the variation in the noise distribution is relatively small.
\end{remark}

Note that the noise term within the
Markov chain, $W$, does not have the same distribution as the noise in the
proposal, $W^*$, since, for example, moves away from positive values of $W$
will be more likely to be rejected than moves away from negative
values of $W$.
In the notation of Section~\ref{sect.psrwm.intro}, since $W^*$ is a
function of $\mathbf{V}$,
$q_{\mathrm{aux}}(\mathbf{x}^*,\mathbf{v})$ now gives rise to $g^*(w^*)$,
the density of the noise in the estimate of the log-target,
which is independent of $\mathbf{x}^*$. Integrating \eqref{eqn.joint.stat.v}
gives the joint stationary density of the
Markov chain $(\mathbf{X},W)$ as
%
%e2.3 #&#
\begin{equation}
\label{eqn.joint.stat.w}
g^*(w)e^w\pi(\mathbf{x}).
\end{equation}
This is Lemma $1$ of \cite{Pittetal2012}.
Under Assumption~\ref{ass.noise.diff.indep}, $W$ and $\mathbf{X}$
are therefore
independent, and the stationary density of $W$ is $g^*(w)e^{w}$.
%how one can approximate this situation in practice. There is a
%discussion in section $3.4$ of \cite{Pittetal2012}]}

%s2.3 #&#
\subsection{High-dimensional target distribution}
\label{sect.highdim}
%now find this and $\pi^{(d)}$ and $s^g_d$ and $s^2(d)$. I don't care
%which we use but it should be consistent!}
We describe in this section conditions on the sequence of target
densities $\pi^{(d)}$
that ensure that the quantity $\log [ \pi^{(d)}(\mathbf{X}^*) /
\pi^{(d)}(\mathbf{X})  ]$ behaves asymptotically
as a Gaussian distribution under an appropriate choice of jump scaling
$\lambda^{(d)}$.
The main\vspace*{0.5pt} assumption is that there exist sequences of scalings $s_{g}^{(d)}>0$
and $s_{L}^{(d)}>0$ for the gradient and the Laplacian of the log-likelihood
$\log  \pi^{(d)}$ such that the following two limits hold in probability:
%
% MAIN CONDITION
%
%e2.4 #&#
\begin{equation}
\label{eq.rescaled.grad.hess}
\hspace*{5pt}\lim_{d \to\infty} \frac{  \| \nabla\log  \pi^{(d)}
(\mathbf{X}^{({d})})  \|}{s_{g}^{(d)}} = 1\quad \mbox{and}\quad \lim
_{d \to\infty} \frac{\Delta  \log  \pi^{(d)}(\mathbf
{X}^{({d})})}{s_{L}^{(d)}} = -1,
\end{equation}
for $\mathbf{X}^{({d})} \stackrel{\mathcal{D}}{\sim}\pi^{(d)}$.
In the rest of this article
we assume that the sequence of densities $\pi^{(d)}$
is such that for each index $i \geq1$, with all components of $\mathbf{x}$ fixed except
the $i$th, the $i$th component satisfies
%
%e2.5 #&#
\begin{equation}
\label{eqn.std.reg}
\frac{\partial\pi^{(d)}}{\partial x_i} \to0 \qquad\mbox{as } \llvert {x_i}
\rrvert \to\infty.
\end{equation}
Under this regularity condition, an integration by parts shows
that
\[
\mathbb{E} \bigl[ \bigl\|\nabla\log\pi^{(d)}\bigl(\mathbf{X}^{({d})}\bigr)\bigr\|^2
 \bigr] = -\mathbb{E} \bigl[ \Delta\log\pi^{(d)}\bigl(\mathbf{X}^{({d})}\bigr) \bigr].
\]
Equation \eqref{eq.rescaled.grad.hess} thus yields $\lim_{d \to
\infty} (s_{g}^{(d)})^2 / s_{L}^{(d)} = 1$.
We will suppose from now on, without loss of generality, that
$s_{g}^{(d)} = \sqrt{s_{L}^{(d)}} =: s^{(d)}$.
We also require that no single component of the local Hessian
$H^{(d)}(\mathbf{x}) :=  [ \partial^2_{ij} \log\pi
^{(d)}(\mathbf{x}) ]_{0 \leq i,j \leq d}$ dominate the others in
the sense
that the limit
%
%e2.6 #&#
\begin{equation}
\label{eqn.eigen.cip}
\lim_{d \to\infty} \frac{\operatorname{Trace} [  (H^{(d)}
)^2(\mathbf{X}^{({d})})  ]}{ ( s^{(d)} )^4} = 0
\end{equation}
holds in probability. We also assume that the Hessian matrix is
sufficiently regular so that for any $\sigma^2, \varepsilon> 0$ and
$\mathbf{Z}^{(d)} \stackrel{\mathcal{D}}{\sim}\mathbf{N}(0,\mathbf
{I}^{(d)})$
%e2.7 #&#
\begin{eqnarray}
\hspace*{2pt}\qquad&& \lim_{d \to\infty} \mathbb{P} \biggl( \sup
_{t \in(0,1)} \biggl| \frac{\langle\mathbf{Z}^{(d)}, [H^{(d)}(\mathbf{X}^{({d})}+t
\sigma  \mathbf{Z}^{(d)}/s^{(d)})-H^{(d)}(\mathbf{X}^{({d})})]
\mathbf{Z}^{(d)} \rangle}{(s^{(d)})^2} \biggr| > \varepsilon \biggr)
\nonumber
\\[-8pt]
\label{eqn.hessian.regularity}
\\[-8pt]
\nonumber
&& \qquad = 0.\hspace*{-12pt}
\end{eqnarray}
These conditions are discussed in
detail in \cite{Sherlock2013} where they are shown to
hold, for example, when the target is the joint distribution of
successive elements of a class of finite-order  multivariate
Markov processes.
The targets\vspace*{1pt} considered in
\cite{RobertsGelmanGilks1997,RobertsRosenthal2001} and
Section~\ref{sec.diff.lim} all satisfy the conditions with $s^{(d)}
\propto d^{1/2}$. We record the conditions formally as:

%as2 #&#
\begin{assumptions}
\label{assump.lim.alpha}
The sequence of densities $\pi^{(d)}$ satisfies equations
\eqref{eq.rescaled.grad.hess}, \eqref{eqn.eigen.cip}, \eqref{eqn.hessian.regularity},
and the regularity condition \eqref{eqn.std.reg}.
\end{assumptions}

We shall show in next section that under these assumptions
the choice of jump size
%
%e2.8 #&#
\begin{equation}
\label{eq.jump.size}
\lambda^{(d)} := \frac{\ell}{s^{(d)}}
\end{equation}
for a parameter $\ell>0$ leads to a Gaussian asymptotic
behaviour for\break  $\log[\pi^{(d)}(\mathbf{X}^*) / \pi^{(d)}(\mathbf{X})]$.
This ensures that for high dimensions, the mean acceptance probability
$\alpha^{(d)}(\ell)$
of the MCMC algorithm,
\[
\alpha^{(d)}(\ell) := \mathbb{E} \biggl[ 1 \wedge\frac{\pi^{(d)} (\mathbf
{X}^{(d)}+\lambda^{(d)} \mathbf{Z}^{(d)}  )e^{W^*}}{\pi
^{(d)} (\mathbf{X}^{(d)} ) e^W}
\biggr],
\]
stays bounded away from zero and one.

%
% DO NOT DELETE THE FOLLWING
%
%The condition [1] show that
%For simplicity, we drop the index $d$. Notice that the quantity $\log
%also reads
%, dt \Big\} Z}.
%Slutsky's lemma shows that it suffices to show that
% \mbox{and}  \bra{Z, \Big\{ \int_{t=0}^2 \Delta(X,X^*,t)
%  (1-t) , dt \Big\} Z} \cip0.
%That is straightforward.

%s2.4 #&#
\subsection{Expected squared jump distance}
\label{sect.setupesjd}
A standard measure of efficiency for local algorithms is
the Euclidian expected squared jumping distance
(e.g., \cite{SherlockRoberts2009,Sherlock2013,BeskosRobertsStuart2009})
usually defined as $\mathbb{E} \| \mathbf{X}_{k+1} - \mathbf{X}_k  \|^2 $.
Consider, for example, a target with elliptical contours, or one which
has components which are
independent and identically distributed up to a scale parameter. In
such situations the
Euclidean ESJD is dominated by those components with a larger scale.
%Optimising Euclidean ESJD would therefore optimise the exploration of
%these
%components at the expense of the other components.
We would prefer an
efficiency criterion which weights components at least approximately
equally, so that moves along each component are considered relative to
the scale of variability of that component. A squared Mahalanobis
distance is
the natural extension of Euclidean ESJD, and in the case of the two
example targets mentioned above, it
is exactly the correct generalisation of Euclidean ESJD.
We therefore define a generalised potential squared jump distance for a
single iteration
with respect to some $d\times d$ positive definite symmetric matrix
$\mathbf{T}^{(d)}$,
$\mathbb{E} [    \| \mathbf{X}^{(d)}_{k+1} - \mathbf
{X}^{(d)}_k  \|_{\mathbf{T}^{(d)}}^2    ]$,
where the Markov chain $\{ \mathbf{X}^{(d)}_k\}_{k \geq0}$ is assumed
to evolve at stationarity
and $\|z\|^2_{\mathbf{T}^{(d)}} := \langle z,\mathbf{T}^{(d)} z
\rangle$.
We will require that, in the limit as $d\rightarrow\infty$, no
one principal component of $\mathbf{T}^{(d)}$ dominates the others in
the sense that
%More formally, let $\tau_{max}^{(d)}$ be the maximum eigenvalue of $
%
%e2.9 #&#
\begin{equation}
\label{eqn.eccentric.distance} %\frac{\tau_{max}^{(d)}}{\Trace{\bmTc{d}}}
\operatorname{Trace} \bigl[{ \bigl(
\mathbf{T}^{({d})} \bigr)^2} \bigr] / \operatorname{Trace}
\bigl[{\mathbf{T}^{({d})}} \bigr]^2 \rightarrow0.
\end{equation}
Clearly, \eqref{eqn.eccentric.distance} is satisfied when $\mathbf
{T}^{({d})} = I_d$ (i.e., Euclidian ESJD).

%
% MAIN THEOREM
%
%th1 #&#
\begin{theorem}
\label{thm.asymp.analysis}
Consider a PsMRWM algorithm. Assume that
the additive noise satisfies
Assumption~\ref{ass.noise.diff.indep}, the sequence of densities
$\pi^{(d)}$ satisfy Assumption~\ref{assump.lim.alpha}, and
the sequence of jump distance matrices $\mathbf{T}^{({d})}$
satisfy~\eqref{eqn.eccentric.distance}.
Assume further that the jump size $\lambda^{(d)}$ is given
by~\eqref{eq.jump.size} for some fixed $\ell>0$.
\begin{longlist}[(2)]
\item[(1)] \textup{Acceptance probability}.
The mean acceptance probabilities $\alpha^{(d)}(\ell)$ converge
as $d\to\infty$ to a nontrivial value $\alpha(\ell)$,
%
%e2.10 #&#
\begin{equation}
\label{eqn.gen.lim.alpha}
\lim_{d \to\infty}{\alpha^{(d)}(\ell)}= 2
\times\mathbb{E} \biggl[ \Phi \biggl(\frac{B}{\ell}-\frac{\ell
}{2} \biggr)
\biggr] =: \alpha(\ell),
\end{equation}
with $B$ as in~\eqref{WBdef}, where $\Phi$ is
the cumulative distribution of a standard Gaussian distribution.
\item[(2)] \textup{Expected squared jump distance}.
A rescaled expected squared jump distance converges
as $d\to\infty$ to a related limit,
%
%e2.11 #&#
\begin{equation}
\label{eqn.gen.lim.esjd}
\lim_{d \to\infty} \frac{  (s^{(d)} )^2}{\operatorname{Trace}
[\mathbf{T}^{({d})}]} \times\mathbb{E}
\bigl\| \mathbf{X}^{(d)}_{k+1} - \mathbf{X}^{(d)}_k
\bigr\|_{\mathbf{T}^{(d)}}^2 = \ell^2 \times \alpha(\ell) =: J(
\ell).
\end{equation}
\end{longlist}
\end{theorem}

Theorem~\ref{thm.asymp.analysis} is proved in
Section~\ref{sec.proof.main.thm}.
%(Since for PsMRWM the process
%$\{ \bmX^{(d)}_k \}_{k \geq0}$ is not itself a Markov chain, the
%mathematical analysis is considerably more intricate.)
It establishes limiting values for the
acceptance probability and expected squared jump distance, and more
importantly for the relationship between them, which is crucial to
establishing optimality results as we shall see. Further,
\eqref{eqn.gen.lim.esjd}
shows that, as is common in scaling problems for MCMC algorithms (e.g.,
in \cite{RobertsGelmanGilks1997,RobertsRosenthal1998}),
the ESJD decomposes into
the product of the acceptance probability $\alpha(\ell)$
and the expected squared \textit{proposed} jumping distance~$\ell^2$,
implying an asymptotic independence between
the size of the proposed move and the acceptance event.
As in the RWM case, we wish
to be able to consider $J(\ell)$ to be a function of the asymptotic acceptance
rate $\alpha(\ell)$. Our next result, which is proved in Section~\ref{sec.proof.bijection.}, shows that this is indeed possible.

% Proposition~\ref{prop.func.of.alpha} below shows that for a given
% distribution of the noise $W^*$, the efficiency of exploration of
% the limiting target distribution using the PsMRWM algorithm can in
% principle be characterised by the limiting mean acceptance probability
% $\alpha(\ell)$ as an alternative to the scaling parameter $\ell$.
%
% The practical implications are
% the same as for the standard RWM algorithm since, in either case,
% $\ell$ is only known up to a multiplicative constant, whereas an
% estimate of $\alpha(\ell)$ can be observed empirically.

%pr1 #&#
\begin{prop}
\label{prop.func.of.alpha}
For a PsMRWM algorithm with noise difference $B$ as in~\eqref{WBdef},
with jump size determined by $\ell>0$ as in~\eqref{eq.jump.size},
and with limiting asymptotic acceptance rate $\alpha(\ell)$
as in~\eqref{eqn.gen.lim.alpha},
the mapping $\ell\mapsto\alpha(\ell)$ is a continuous decreasing
bijection from $(0,+\infty)$ to $(0,\alpha_{\max}]$,
where
\[
\alpha_{\max} := \lim_{\ell\to0} \alpha(\ell) = 2 \times
\mathbb{P}[B>0].
\]
\end{prop}

% Now, in real scenarios our assumptions, and in particular Assumptions
% \ref{ass.noise.diff.indep} that the distribution of the noise
% difference is independent of the state, will not hold precisely and
% could affect the choice of tuning strategy. An alternative to altering
% the scaling to aim for a particular acceptance rate would be to vary
% the scaling in order to optimise some function such as the empirical
% ESJD or integrated autocorrelation time.

Proposition~\ref{prop.func.of.alpha} yields that $\alpha_{\max} = \sup_{\ell> 0}   \alpha(\ell)$.
When there is no noise in the estimate of the target, as already proved
in \cite{RobertsGelmanGilks1997}, the acceptance
rate simplifies to $\alpha_0(\ell) := 2   \Phi(-\ell/2)$, and the
associated expected squared jump distance reads $J_0(\ell) = \ell^2
  \alpha_0(\ell)$. Thus we may also consider the asymptotic
efficiency of a pseudo-marginal algorithm relative to the idealised
algorithm if the target were known precisely by defining $J_{\mathrm{rel}}(\ell
) = J(\ell) / J_0(\ell)$, which also reads
%e2.12 #&#
\begin{equation}
\label{eqn.rel.eff.gen}
J_{\mathrm{rel}}(\ell)= \frac{1}{\Phi (-\ell/ 2 )}\mathbb{E} \biggl[{\Phi
\biggl(\frac{B}{\ell}-\frac{\ell}{2} \biggr)} \biggr].% \big/ \Phi\left(-
\end{equation}

The following proposition, which is proved in Section~\ref{proof.prop.bds}, shows that the
relative efficiency can never exceed unity and that it is bounded
below by the acceptance rate in the limit as $\ell\rightarrow0$.

%pr2 #&#
\begin{prop}
\label{prop.rel.eff.bds}
With $\alpha(\ell)$ and $J_{\mathrm{rel}}(\ell)$ as defined in
\eqref{eqn.gen.lim.alpha} and \eqref{eqn.rel.eff.gen} respectively,
\[
\alpha_{\max} \le J_{\mathrm{rel}}(\ell) \le 1.
\]
\end{prop}

%f1
%f1 #&#
\begin{figure}[t]

\includegraphics{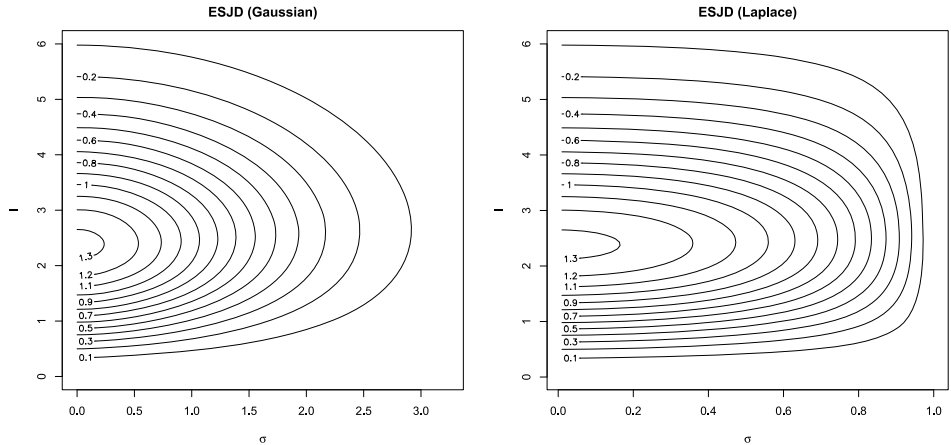}

\caption{Contour plots of the asymptotic expected squared jump distance
$J(\ell)$ from \protect\eqref{eqn.gen.lim.esjd}
plotted as a function of the
scaling parameter $\ell$
and of the standard deviation, $\sigma$, of the additive noise. In the
left-hand panel
the additive noise in the log-target is assumed to be Gaussian,
% ($W^*\sim\Normal(-\sigma^2/2,\sigma^2)$)
and in right-hand panel it is assumed to
have a Laplace distribution.}
%(with mean $log(1-\sigma^2/2)$ and scale parameter
%$\sigma/\sqrt{2}$)
% The bottom figures correspond to
% multiplicative Gamma noise \eqref{eqn.noise.one.gamma} in the target.
%, and in this case the variance of the noise in the log-target is
% estimated from a very large Monte Carlo sample.
\label{fig.esjd.acc}
\end{figure}

\begin{figure}

\includegraphics{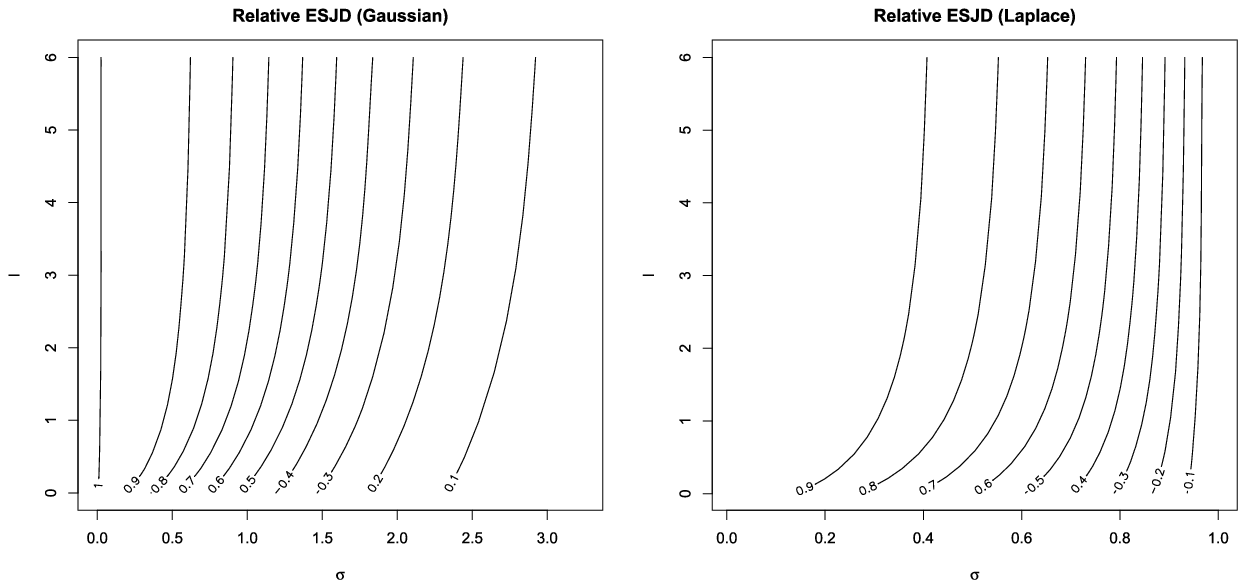}

\caption{Contour plots of $J_{\mathrm{rel}}(\ell)$ from \protect\eqref{eqn.rel.eff.gen}, the
asymptotic expected squared jump distance relative to the idealised algorithm,
plotted as a function of the
scaling parameter $\ell$
and of the standard deviation, $\sigma$, of the additive noise. In the
left-hand panel the additive noise in the log-target is assumed to be Gaussian,
% ($W^*\sim\Normal(-\sigma^2/2,\sigma^2)$)
and in the right-hand panel it is assumed to
have a Laplace distribution.}
%(with mean $log(1-\sigma^2/2)$ and scale parameter
%$\sigma/\sqrt{2}$)
% The bottom figures correspond to
% multiplicative Gamma noise \eqref{eqn.noise.one.gamma} in the target.
%, and in this case the variance of the noise in the log-target is
% estimated from a very large Monte Carlo sample.
\label{fig.esjd.rel.acc}
\end{figure}

The quantities $\alpha(\ell)$, $J(\ell)$ and $J_{\mathrm{rel}}(\ell
)$ depend upon
the distribution of $B$, and hence on the distribution of the
additive noise $W$ from~\eqref{WBdef}.
Figure~\ref{fig.esjd.acc}
considers two particular cases: where the distribution of the
additive noise is Gaussian, that is,
$W^*\sim\mathbf{N}(-\sigma^2/2,\sigma^2)$
(which we shall consider further in
Section~\ref{sect.optimising}), and where the distribution of the
additive noise is Laplace (i.e., double-exponential),
with mean $\log(1-\sigma^2/2)$ and scale parameter $\sigma/\sqrt{2}$.
For each of these two cases, it shows a \textit{contour plot} of
$J(\ell)$ as a function of the proposal scaling
parameter $\ell$ and of the standard deviation of the additive noise,
$\sigma$. Figure~\ref{fig.esjd.rel.acc} shows the equivalent plots
for $J_{\mathrm{rel}}(\ell)$.

%f2
%f2 #&#

Our ultimate goal is often to choose $\ell$ to \textit{maximise}
$J(\ell)$, and thus obtain an \textit{optimal} limiting diffusion
(and hence an approximately optimal algorithm for finite $d$ too).
We shall use Theorem~\ref{thm.asymp.analysis}
to establish an optimal acceptance rate
in a particular limiting
%high-dimensional
regime, in Section~\ref{sect.opt.std} below.

Figure~\ref{fig.esjd.rel.acc} illustrates that, except for small
values of the
scaling, the relative efficiency for a given noise distribution is
relatively insensitive to the scaling. Related to this, from Figure~\ref{fig.esjd.acc} it appears that the optimal scaling [i.e.,
the value $\ell$ which maximises $J(\ell)$] is relatively insensitive
to the variance of the additive noise. When there is no noise, the
optimum is $\hat{\ell}_0\approx2.38$ as first noted in
\cite{RobertsGelmanGilks1997}; however, the optimum remains close to
$2.5$
across a range of variances for both choices of noise distribution.

% The choice of scaling strategy will therefore depend on the particular
% problem.

For these two examples, as might be expected, for
any given scaling of the random walk proposal, the
efficiency relative to the idealised algorithm decreases as the
standard deviation of the noise increases, a phenomenon that is
investigated more generally in \cite{AndrieuVihola2014}.
Thus there is an implicit
\textit{cost} of having to estimate the target density. As a result of
this, we should not expect the optimal acceptance probability for RWM
of $0.234$ to hold here.

%s2.5 #&#
\subsection{Diffusion limit}
\label{sec.diff.lim}

We next prove that PsMRWM in high dimensions can be well-approximated
by an appropriate diffusion limit (obtained as $d \to\infty$).
This provides further justification for measuring efficiency by
the ESJD, as discussed in detail in \cite{RobRos12}. Briefly,
the limiting ESJD (suitably scaled) is equal to the square of the
limiting process's diffusion coefficient, $h$ say. By a simple time
change argument, the asymptotic variance of \textit{any} Monte Carlo
estimate of interest is inversely proportional to $h$. Minimising
variance is thus equivalent to maximising $h$; that is, $h$ becomes (at
least in the limit) unambiguously the right quantity to optimise.
By constrast, MCMC algorithms which have nondiffusion limits can
behave in very different ways, and ESJD may not be an appropriate
way to compare algorithms in such cases.

We shall consider in this section the PsMRWM algorithm applied to a
sequence of simple i.i.d. target densities
\[
\pi^{(d)}(x_1, \ldots, x_d) = \prod
_{i=1}^d f(x_i),
\]
where $f$ is a one-dimensional probability density. We assume
throughout this section
that the following regularity assumptions hold.
%These are mainly technical assumptions and we believe that they could
%be relaxed at the expense of considerably more intricate proofs.
%
% ASSUMPTION on f

%as3 #&#
\begin{assumptions} \label{assump.f}
The first four moments of the distribution with density $f$
are finite.
The log-likelihood mapping
$x \mapsto\log f(x)$ is smooth with second, third, and fourth
derivatives globally bounded.
\end{assumptions}

One can verify that under Assumption~\ref{assump.f}, the target
$\pi^{(d)}$ satisfies Assumption~\ref{assump.lim.alpha}. It is
important to stress that the $\mathrm{ESJD}$ analysis of Section~\ref{sect.setupesjd} only relies on the weaker
Assumption~\ref{assump.lim.alpha}, and as discussed at the end of the previous
section, is valid for much more general target distributions than the
ones with i.i.d. coordinates considered in this section. The stronger
Assumption~\ref{assump.f} are standard in the diffusion-limit
literature and are, perhaps, the simplest from which a diffusion limit
is expected to result \cite{RobertsGelmanGilks1997}. However, these
i.i.d. assumptions have been relaxed in various directions \cite{breyer2004optimal,BedardA2007,BedardRosenthal2008,BeskosRobertsStuart2009,PST12},
and we believe that our diffusion limit Theorem~\ref{thm.diff.lim}
could also be extended to similar settings at the cost of considerably
less transparent proofs.

In the remainder of this article we consider the sequences of scaling
functions $\sqrt{s_L^{(d)}}=s_g^{(d)}:=\sqrt{I \times d}$, with
%
%e2.13 #&#
\begin{equation}
\label{e.I}
I:= \mathbb{E} \bigl[ \bigl\{ \bigl( \log f(X)
\bigr)' \bigr\}^2 \bigr] = -\mathbb{E} \bigl[ \bigl(
\log f(X) \bigr)^{\prime\prime} \bigr]
\end{equation}
and $X \stackrel{\mathcal{D}}{\sim}f(x)   \,dx$. Indeed, equation
\eqref{eq.rescaled.grad.hess} is satisfied; consequently,
%we define $s^{(d)}=\sqrt{I \times d}$ and
for a tuning parameter $\ell>0$, we consider $d$-dimensional RWM
proposals with scaling
%
%e2.14 #&#
\begin{equation}
\label{eq.jump.size.diff.lim} \lambda^{(d)} := \ell I^{-1/2}
\delta^{1/2} \qquad\mbox{with } \delta= 1/d
\end{equation}
as in~\eqref{eq.jump.size}. The quantity $I$, which quantifies the
roughness and the scale of the marginal density $f(x)   \,dx$, has been
introduced in the definition of the RWM jump-size \eqref{eq.jump.size.diff.lim}
so that all our limiting results on the \textit{optimal} choice of parameter $\ell$ are independent of $f(x)   \,dx$.
The main\vspace*{1pt} result of this section is a diffusion
limit for a rescaled version $V^{(d)}$ of the first coordinate process.
For time $t \geq0$ we define
the piecewise-constant continuous-time process
\[
V^{(d)}(t) := X^{(d)}_{\lfloor{dt} \rfloor,1}
\]
with the notation $\mathbf{X}^{(d)}_k = (X^{(d)}_{k,1}, \ldots,
X^{(d)}_{k,d} ) \in\mathbb{R}^d$ so that $V^{(d)}(t)$ is the first
coordinate of $\mathbf{X}^{(d)}_{\lfloor{dt} \rfloor}$.
Note that in general the process $V^{(d)}$ is not Markovian.
The next theorem shows that nevertheless,
in the limit $d \to\infty$, the process $V^{(d)}$
converges weakly to an explicit Langevin diffusion.
This result thus generalises the original RWM diffusion limit proved
in \cite{RobertsGelmanGilks1997}.

%th2 #&#
\begin{theorem} \label{thm.diff.lim}
Let $T>0$ be a finite time horizon.
For all $d\geq1$ let each Markov chain and the additive noise satisfy
Assumption~\ref{ass.noise.diff.indep}, let the sequence
of product form densities $\pi^{(d)}$ satisfy the regularity
Assumption~\ref{assump.f} and set the scale of the jump proposals as
in equation \eqref{eq.jump.size.diff.lim}.
Then, as $d \to\infty$,
\[
V^{(d)} \Rightarrow V
\]
in the Skorokhod topology on $D([0,T])$, where $V$ satisfies the Langevin
SDE
%
%e2.15 #&#
\begin{equation}
\label{e.limiting.diffusion}
dV_t = h^{1/2}(\ell) \, dB_t +
\tfrac{1}{2} h(\ell) \nabla\log f(V_t) \,dt
\end{equation}
with initial distribution $V_0 \stackrel{\mathcal{D}}{\sim}f$ and
$B_t$ a standard Brownian motion. The speed function
$h$ is proportional to the asymptotic rescaled ESJD function $J$,
\[
h(\ell) = J(\ell) / I,
\]
with the constant of proportionality $I$ defined by equation \eqref{e.I}.
\end{theorem}

The time change argument discussed before Theorem~\ref{thm.diff.lim}
shows that the quantity $J_{\mathrm{rel}}$ exactly measure the loss of
mixing efficiency (computational time not taken into consideration)
when exact evaluations of the target density are replaced by unbiased estimates;
as already mentioned, the pseudo-marginal algorithm always has worse
mixing properties than the idealised algorithm.

%s3 #&#
\section{Optimising the PsMRWM}
\label{sect.optimising}

We next consider the question of optimising the PsMRWM.
Now, when examining the efficiency of a standard RWM,
the expected computation (CPU) time is usually
not taken into account since it is implicitly assumed to be independent of
the choice of tuning parameter(s).
This may indeed be approximately true for the RWM. However, for the
PsMRWM the
expected CPU time for a single iteration of the algorithm is
usually approximately inversely proportional to the variance of the
estimator $\hat{\pi}(x)$.
For this reason, we measure the efficiency of the PsMRWM through a
rescaled version of the ESJD,
%
%e3.1 #&#
\begin{equation}
\label{def.efficiency}
(\mathrm{Efficiency}) := \frac{(\mathrm{Expected} \  \mathrm{Square} \ \mathrm{Jump} \
\mathrm{Distance})}{(\mathrm{Expected} \  \mathrm{one\mbox{-}step} \ \mathrm{computing} \ \mathrm{time})}.
\end{equation}
Of course, for any increasing function $F$, the quantity
$F\mbox{(ESJD)}/\mbox{(Expected}\break \mbox{one-step computing time)}$ is a
possible measure of efficiency. However, the discussion at the start
of Section~\ref{sec.diff.lim} indicates that \eqref{def.efficiency}
is the appropriate measure of efficiency in the high-dimensional
asymptotic regime considered in this article.

In the remainder of this section, we implicitly assume that the target
distributions satisfy Assumption~\ref{assump.lim.alpha}.

%s3.1 #&#
\subsection{Standard (Gaussian) regime}
\label{sec.standard.regime}
We shall restrict attention to the case in which the
additive noise follows a Gaussian distribution. More precisely, we
shall assume the following, which we shall refer to for brevity as
``the standard asymptotic regime'' (SAR):
%
% SAR Assumption

%as4 #&#
\begin{assumptions}
\label{ass.standard.regime}
For each $x \in\mathcal{X}$ and $\sigma^2 > 0$, we have an unbiased
estimator $\hat{\pi}(x)$ of $\pi(x)$, such that
$\log\hat{\pi}(x)$ follows a Gaussian distribution with
variance~$\sigma^2$. Furthermore, the expected one-step computing time
is inversely proportional to~$\sigma^2$.
\end{assumptions}

Intuitively, Assumption~\ref{ass.standard.regime} are designed to
model the situation where $\pi(x)$ is estimated as a product of $n$
averages of $m$ i.i.d. samples in the limit as $n\rightarrow\infty$
and with $m\propto n$. For a fixed large $n$, approximate normality
follows from the central limit theorem; moreover $\sigma^2 \approx
c/m$ for some
$c>0$, and the computational time is proportional to $m$ and hence to
$1/\sigma^2$. Assumption~\ref{ass.standard.regime}
have recently been shown to hold more generally, in the context of
particle filtering for a hidden
Markov model; see \cite{berard2013lognormal}.
There are other natural situations where multiplicative forms for the
importance sampling estimator of the likelihood might make the estimator
well-approximated as a log-Gaussian, for example, in correcting for a PAC
likelihood approximation; see \cite{LiSte03}.

Under the SAR of
Assumption~\ref{ass.standard.regime},
we will prove an optimality result
in Section~\ref{sect.opt.std}
which specifies a particular optimal variance for the estimate of the
log-target.

% The proof assumes that the estimate is Gaussian and that its
% variance is inversely proportional to the computational effort
% involved in creating it; for importance sampling the computational
% effort is proportional to the number of unbiased estimates and
% for the particle filter it is proportional to the number $m$ of
% particles. The Gaussianity of the log-target arises naturally,
% for example, from a `large $T$, fixed $m$' scenario or, via Taylor
% expansion, from a `fixed $T$, large $m$' scenario. However with the
% former the variance of the log-target grows without bound and is
% not proportional to $1/m$; with the latter, while the variance is
% (eventually) proportional to the reciprocal of $m$, it shrinks to
% zero as $m \rightarrow\infty$. To create a Gaussian log-target that
% requires no rescaling for its variance to be finite and non-zero,
% $m$ and $T$ must increase in tandem.

%s3.2 #&#
\subsection{Optimisation under the standard asymptotic regime}
\label{sect.opt.std}

In this section we consider a sequence $\pi^{(d)}$ of target
distributions satisfying Assumption~\ref{assump.lim.alpha} and assume
that each unbiased estimator satisfies the independence in
Assumption~\ref{ass.noise.diff.indep}. Under these assumptions, the rescaled ESJD
of the PsMRWM algorithm with jump size \eqref{eq.jump.size} is described
by Theorem~\ref{thm.asymp.analysis}. Under the SAR, that is,
Assumption~\ref{ass.standard.regime}, and with
$\operatorname{Var}  [\log\hat{\pi}(x) ] = \sigma^2$,
the noise
difference is $B \stackrel{\mathcal{D}}{\sim}\mathbf{N}(-\sigma
^2,2\sigma^2)$. Since the mean
one-step computing time is assumed to be inversely proportional to
the variance, $\sigma^2$, the asymptotic efficiency, as $d \to\infty$,
is proportional to
%
%e3.2 #&#
\begin{equation}
\label{def.efficiency.rescaled} % \mbox{(Efficiency)}   \propto
\sigma^2 \times J_{\sigma^2}(\ell)
=: \mathbf{Eff}_{\sigma^2}(\ell),
\end{equation}
where $J_{\sigma^2}(\ell)$
stands for the asymptotic rescaled $\mbox{ESJD}$ identified
in Theorem~\ref{thm.asymp.analysis}, that is, $J(\ell)$,
in the special case where $B \stackrel{\mathcal{D}}{\sim}\mathbf
{N}(-\sigma^2,2\sigma^2)$.

\begin{figure}

\includegraphics{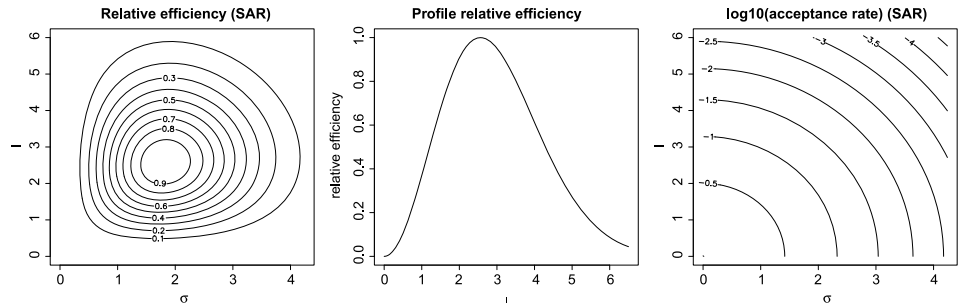}

\caption{Contour plots of the theoretical relative efficiency
$\mathbf{Eff}_{\sigma^2}(\ell)/\mathbf{Eff}_{\sigma^2_{\mathrm{opt}}}(l_{\mathrm{opt}})$,
and of the base-10 logarithm
of the asymptotic acceptance probability $\alpha(\ell)$, and a plot of
the profile relative efficiency
$\mathbf{Eff}_{\sigma^2_{\mathrm{opt}}(\ell)}(\ell)/\mathbf{Eff}_{\sigma
^2_{\mathrm{opt}}}(l_{\mathrm{opt}})$,
all for the scenario where the additive noise arises from the SAR.}
% assuming additive Gaussian noise in the
% estimate of the log-target ($W^*\sim\Normal(-\sigma^2/2,\sigma^2)$).
\label{fig.Gauss.eff.acc}
\end{figure}

Figure~\ref{fig.Gauss.eff.acc} provides a contour plot of this
efficiency $\mathbf{Eff}_{\sigma^2}(\ell)$, relative to the highest
achievable
efficiency, and of the logarithm of the asymptotic acceptance rate
$\alpha(\ell)$, both as
functions of the scaling parameter $\ell$ and of the standard deviation,
$\sigma$. It also\vspace*{-1.5pt} provides
a plot of the profile $\mathbf{Eff}_{\sigma^2_{\mathrm{opt}}(\ell)}(\ell)$
as a
function of $\ell$, again relative to the highest achievable value.

As previously suggested by Figure~\ref{fig.esjd.acc},
we see that the conditional optimal value of $\ell$ is relatively
insensitive to the value of $\sigma$.

The point at which the maximal efficiency is achieved is
detailed precisely in Corollary~\ref{cor.max.Gauss} below.

% Figure~\ref{fig.Gauss.eff.acc} shows that the converse is also true.

%f3
%f3 #&#

%co1 #&#
\begin{cor}
\label{cor.max.Gauss}
The\vspace*{1pt} efficiency $\mathbf{Eff}_{\sigma^2}(\ell)$
is maximised (to three decimal places) when the variance $\sigma^2$
of the log-noise is
\[
\sigma_{\mathrm{opt}}^2  =  3.283,
\]
and the scaling parameter $\ell$ is
\[
\ell_{\mathrm{opt}}  =  2.562,
\]
at which point the corresponding asymptotic acceptance rate is
\[
\alpha_{\mathrm{opt}} = 7.001\%.
\]
As $\sigma^2\rightarrow\infty$ the optimal scaling satisfies
$\ell_{\mathrm{opt}}(\sigma)\rightarrow 2 \sqrt{2}$, and as $\ell\rightarrow
\infty$ the optimal variance satisfies $\sigma_{\mathrm{opt}}^2(\ell
)\rightarrow 4$.
\end{cor}

\begin{pf}
For convenience, write $\tau^2:=2\sigma^2$, and introduce three
independent standard Gaussian random variables $U,V,Z \stackrel{\mathcal{D}}{\sim}\mathbf{N}(0,1)$. Notice that
$B \stackrel{\mathcal{D}}{\sim}-\tau^2/2 + \tau  U$ and
%
%e3.3 #&#
\begin{eqnarray}
\mathbf{Eff}_{\sigma^2}(\ell) & = &
\tau^2 \ell^2 \mathbb{E} \bigl[ \Phi(B/\ell-\ell/2)
\bigr] \nonumber\\
&=& \tau^2 \ell^2 \mathbb{P} \bigl[V < \bigl(-
\tau^2/2 + \tau U \bigr)/\ell-\ell/2 \bigr]
\nonumber\\
\label{eq.closed.form.eff}&=&  \tau^2 \ell^2 \mathbb{P}\bigl(\ell V - \tau U < - \bigl(
\tau ^2 +\ell^2 \bigr) / 2 \bigr]
\\
\nonumber
&=&  \tau^2
\ell^2 \mathbb{P} \bigl[ \sqrt{\ell^2+\tau^2}
Z < - \bigl(\tau^2 +\ell^2 \bigr) / 2 \bigr]
\\
&= & \tau^2 \ell^2 \Phi \bigl(-\tfrac{1}{2}\sqrt{
\tau^2+\ell ^2} \bigr).\nonumber
\end{eqnarray}
For fixed $\tau^2+\ell^2$, the quantity $\tau^2\ell^2$ is maximised when
$\tau^2=\ell^2$, at which point the efficiency is
$\tau^4\Phi(-\tau/\sqrt{2}) \propto\sigma^4\Phi(-\sigma)$.
%So, the optimal value of $\sigma^2$ is the one which maximises
%the function $\sigma^4\Phi(-\sigma)$.
This is maximised numerically when
$\sigma^2 = \sigma_{\mathrm{opt}}^2 = 3.283$ (to three decimal
places), and at this
point $\ell_{\mathrm{opt}}=\sigma_{\mathrm{opt}} \sqrt{2}$
and $\alpha_{\mathrm{opt}}=2\Phi(-\sigma_{\mathrm{opt}})$ with the
corresponding
numerical values as stated.

Differentiating \eqref{eq.closed.form.eff} with respect to $\ell$ we
find that the optimal scaling satisfies
\[
\Phi \bigl(-\tfrac{1}{2}\sqrt{\ell^2+\tau^2}
\bigr)=\tfrac
{1}{4}\ell^2\varphi \bigl(-\tfrac{1}{2}
\sqrt{\ell^2+\tau^2} \bigr) / \sqrt{\ell^2 +
\tau^2}.
\]
The result for large $\tau^2$ follows from the relationship
$\Phi(-x)\sim\varphi(x)/x$ as $x \to\infty$. The symmetry of the
function $(\ell^2,\tau^2) \mapsto\mathbf{Eff}_{\sigma^2}(\ell)$
in $\tau$ and
$\ell$ then provides the result for large $\ell$.
\end{pf}

%If the variance is approximately optimal, $\Var{W^*_{n,m}}\approx3.3$,
%then since $W^*_{n,m}$ is the sum of $n$ independent terms, at least
%one of these must have a variance of at least $3.3/n$. If $n$ is small
%then this contradicts (\ref{eqn.small.tausq}) and the standard
%asymptotic regime does not apply. Therefore, when $n$ is small we may
%$3.3$. \end{remark}

\begin{remarks*}
(1)
This leads to a new optimal scaling for standard Gaussian targets
of $\lambda\approx
\ell_{\mathrm{opt}}/\sqrt{d}$ with $\ell_{\mathrm{opt}}\approx
2.562$, and contrasts with the corresponding formula
$\hat{\ell}_0/\sqrt{d}$, with $\hat{\ell}_0 \approx2.38$, for the
usual random walk Metropolis algorithm \cite{RobertsGelmanGilks1997}; recall that $\hat{\ell}_0$ satisfies
$\hat{\ell}_0 = \mbox{argmin}_{\ell>0}   \ell^2   \Phi(-\ell
/ 2)$.

(2)
In the discussion of Figure~\ref{fig.esjd.acc} it was noted that for a
Gaussian or
Laplace noise
regime the optimal scaling at a particular noise variance, $\sigma^2$, is
insensitive to the value of $\sigma^2$. From Figure~\ref{fig.Gauss.eff.acc} and from the symmetry of expression
\eqref{eq.closed.form.eff}, the optimal variance at a particular
scaling $\ell$ is also insensitive to the value of $\ell$. Moreover
as $\ell\to0$ the optimal
variance is $\hat{\ell}_0^2/2\approx2.83$, which corresponds (at
least to 2 decimal places) with the value obtained in \cite{Doucetetal2014}.

(3) In practice, $\sigma^2$ might be a function of a
\textit{discrete} number $m$ of samples or particles and hence only
take a discrete set of values. In
particular, if the variance in the noise using $m=1$
is already lower than $3.283$, then there can be little
gain in increasing $m$.

(4)
In many problems the computational cost of obtaining an unbiased
estimate of the target is much
larger than the cost of the remainder of the algorithm, but this is
not always the case. Consider therefore
the more general problem where the cost of obtaining a
single unbiased estimate is $t_{\mathrm{rat}}$ times the cost of the remainder
of the algorithm. In this case the efficiency functional should be expressed
as $(\mbox{Efficiency}) = J_{\sigma^2}(\ell)/(1+t_{\mathrm{rat}} \sigma^{-2})$
and the optimal acceptance rate is a function of
$t_{\mathrm{rat}}$ which varies between  $7.0\%$ (as $t_{\mathrm{rat}}\rightarrow
\infty$)
and $23.4\%$ (as $t_{\mathrm{rat}}\rightarrow0$).
\end{remarks*}

Figure~\ref{fig.Gauss.eff.acc} shows that in contrast to the
insensitivity of the optimal scaling to the variance of the noise, the
acceptance rate at this optimum could potentially
vary by a factor of $3$ or more. Thus if a particular scaling of the
jump proposals maximises $J(\ell)$ for some
particular noise distribution and variance, then that scaling should
be close to optimal across a wide range of noise distributions and
variances. However, tuning to a particular acceptance rate,
whilst more straightforward in practice, could lead to a sub-optimal
scaling if the noise distributions encountered in the tuning runs
are not entirely representative of the distributions that will be
encountered during the main run.

Our theory applies in the limit when the dimension $d$ of the (marginal)
target $\mathbf{X}$ goes to infinity. However,
using a similar argument to that in \cite{SherlockRoberts2009}, when
$\mathbf{X}\sim\mathbf{N}(0,\mathbf{I}_d)$, it
can be shown that under the SAR with the proposal as in \eqref{eqn.rwm.prop}
the $\mbox{ESJD}$ and acceptance rate are
\begin{eqnarray*}
\mbox{ESJD}(\lambda,d) &=&  2 \lambda^2 \mathbb{E} \biggl[ \|
\mathbf{Z}\|^2 \Phi \biggl(-\frac{\lambda}{2}\| \mathbf{Z}\| +
\frac{B}{\lambda\|
\mathbf{Z}\|} \biggr) \biggr] \quad\mbox{and}\\
 \alpha(\lambda,d) &=&  \mathbb{E}
\biggl[ \Phi \biggl(-\frac{\lambda}{2}\| \mathbf{Z}\| + \frac{B}{\lambda\|\mathbf{Z}\|}
\biggr) \biggr],
\end{eqnarray*}
where $\mathbf{Z}\stackrel{\mathcal{D}}{\sim}\mathbf{N}(0, \mathbf
{I}_d)$ and $B \stackrel{\mathcal{D}}{\sim}\mathbf{N}(-\sigma^2,
2 \sigma^2)$. Numerical optimisation of the efficiency
function, $\sigma^2 \times\mbox{ESJD}(\lambda,d)$ for
$d=1,2,3,5$, and $10$ produces a steady decrease in
$\hat{\ell}=\hat{\lambda}\sqrt{d}$ from $2.59$ to $2.57$ and in
$\hat{\alpha}$ from $11.5\%$ to $7.7\%$, and a similarly steady
increase in $\hat{\sigma}^2$ from $3.23$ to $3.27$.
Thus, at least for Gaussian targets and with efficiency measured by
ESJD, the asymptotic
results for the optimal scaling and optimal
variance are applicable in any dimension but there may be a small
increase in the optimal acceptance rate, as is found for the
nonpseudo-marginal RWM (e.g., \cite{RobertsRosenthal2001,SherlockRoberts2009}).

In the simulation study of Section~\ref{sect.sim.study} below, we
find that Corollary~\ref{cor.max.Gauss} and its associated formulae
provide a good description of the optimal settings for a particle
filter with $T=50$ and $d=5$.

\section{Simulation study}
\label{sect.sim.study}

In this section we restrict attention to the
%standard asymptotic regime
SAR of Section~\ref{sec.standard.regime}.
% ($n\gg1$ in \eqref{eqn.gen.error.form}).
Corollary~\ref{cor.max.Gauss}
%The optimality result for this regime
%applies in the limit as
%$d\rightarrow\infty$, and provided the distribution of the noise,
%$W^*$, is
%independent of $\bmX$ and has a Gaussian distribution.
%It
suggests that the
optimal efficiency should be
obtained by choosing the number of unbiased estimates, $m$, such that
the variance in the log-target is approximately $3.3$. The scale
parameter, $\lambda$, should be set so that the acceptance rate is
approximately $7\%$. Since the constant of proportionality
relating $\lambda$ and $\ell$ is unknown in practice, we cannot
simply set
$\ell\approx2.56$.

In practice the assumptions underlying this result may not hold: the
dimension of the parameter space is finite,
the distribution of the noise, $W^*$, may not be Gaussian,
and it is likely to also vary with position, $\mathbf{x}^*$. We
conduct a
simulation study to provide an indication of both the extent of and
the effect of such deviations.

We use the Particle Marginal RWM algorithm (PMRWM) of
\cite{AndrieuDoucetHolenstein2010} to perform exact inference for the
Lotka--Volterra predator-prey model;
see
\cite{GolightlyWilkinson2011} for a more detailed description of the
PMRWM
which focusses on this particular class of applications.
Starting from an initial value,
which is, for
simplicity, assumed known, the two-dimensional latent variable $\mathbf{U}$ evolves
according to a Markov jump process (MJP). Each component is observed
at regular intervals with Gaussian
error of an unknown variance. Appendix~\ref{sec.lotka.details}
provides details of the
observation regime and of the
transitions of the MJP and their associated rates. It also provides
the parameter values, the priors and the lengths of the MCMC runs.

%Given a prior distribution $\pi_0(\bmx)$,
%a simple unbiased estimate of the posterior density for $x$ is
%(up to a fixed constant of proportionality)
%where $\bmv(t)$ represents a MJP that has been simulated from the
%initial value $\bmv(0)$ using rate parameters $(x_1,x_2,x_3)$. The
%variance of such an estimate, or by the average of any reasonable
%number, $m$, of such estimates is generally too large to obtain
%reasonable mixing of the Monte Carlo Markov chain and so we employ the
%Particle Marginal RWM algorithm of

An initial run provided an estimate of a central value, $\hat{\mathbf{x}}$
(the vector of posterior medians), and the posterior variance matrix,
$\widehat{\operatorname{Var}}(\mathbf{X})$. Since the shape of the target
distribution, and hence the optimal shape of
the proposal, is unknown, we follow the frequently used strategy for
the RWM (e.g., \cite{SherlockFearnheadRoberts2010}) of setting the
proposal covariance matrix to be proportional to
$\widehat{\operatorname{Var}}(\mathbf{X})$.
%the estimate of the
%target covariance matrix from the
%preliminary run.
From Remark~\ref{rem1} following Corollary~\ref{cor.max.Gauss}, we set
$\mathbf{V}_{\mathrm{prop}}=\gamma^2\times(2.56^2/d)\times\widehat
{\operatorname{Var}}(\mathbf{X})$ with
$\gamma=1$ corresponding to an optimal tuning for a Gaussian target.

Let $\mathbb{M}:=\{50,80,100,150,200,300,400\}$ define the set of choices
for the number of particles, $m$, and let
$\mathbb{G}:=\{0.4,0.6,0.8,1.0,1.2,1.4,1.6\}$ define the set of
choices for the relative scaling, $\gamma$. For each $(m,\gamma)$ in
$\mathbb{M}\times\mathbb{G}$ an MCMC run of at least $2.5\times10^5$
iterations was performed starting from $\hat{\mathbf{x}}$. For
diagnostic purposes runs of at
least $10^4$ iterations were performed with $m\in\mathbb{M}$ and
$\gamma=0$ (so $\mathbf{x}=\hat{\mathbf{x}}$ throughout).

\begin{figure}[t]

\includegraphics{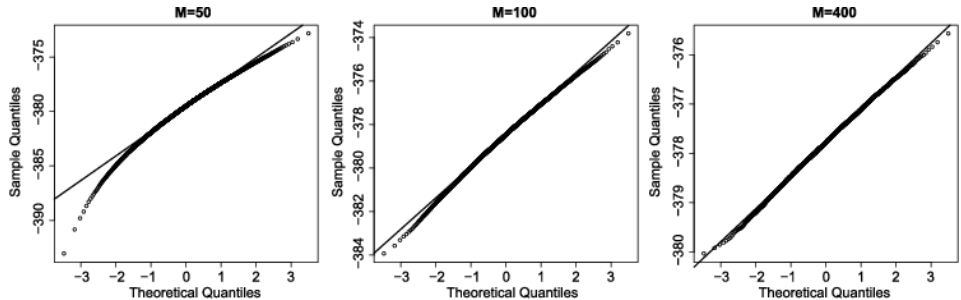}

\caption{Normal QQplots of the noise in the estimate of the log-target at the
a proposed value of the posterior median, $\hat{\mathbf{x}}$, when $m=50$
(left panel), $m=100$ (centre), and
$m=400$ (right).}
\label{fig.QQplots}
\end{figure}

\begin{figure}[b]

\includegraphics{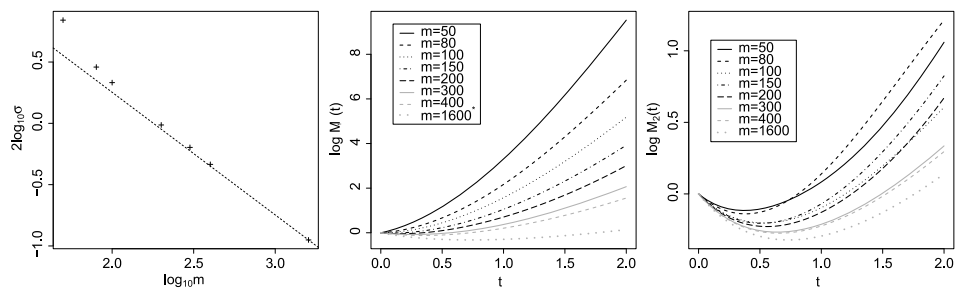}

\caption{In the left panel the logarithm of the empirical variance
of the noise in the estimate of the log-proposal
sampled at $\mathbf{x}=\hat{\mathbf{x}}$ is plotted against the
logarithm of the number of
particles used; the centre and right panels are plots of the
logarithms of the empirical estimates of the moment generating
functions of $\hat{L}$ and $L$ ($M_1(t)$ and $M_2(t)$, resp.)
against $t$. The additional lowest curve in the centre
panel ${}^*$ and in the right-hand panel is the logarithm of
$M_2(t)$ with $m=1600$, and constitutes our best estimate of ``truth.''}
\label{fig.morediags}
\end{figure}

%or less a verification that a CLT holds for the particle filter
%estimate, which is indeed known (but maybe the LV model does not
%satisfy the usual assumption). We could also cite some of the relevant
%literature; I can do that]}
We perform three checks on our assumptions. The diagnostic runs
provide a sample from the distribution of $W^*$, the estimate of the log-target
at a proposed value; this allows us to investigate the second part of
Assumption~\ref{ass.noise.diff.indep} and both parts of Assumption~\ref{ass.standard.regime}. We first examine the SAR
Assumption~\ref{ass.standard.regime}. Figure~\ref{fig.QQplots} shows QQplots for
$m=50$, $m=100$
and $m=400$ against a Gaussian distribution; it is clear that at
$m=50$ the right-hand tail is slightly too
light and the left-hand tail is much heavier than that of a
Gaussian. Similar but much smaller discrepancies are present at
$m=100$, whilst at $m=400$ the noise distribution is almost
indistinguishable from that of a Gaussian.
%f4
%f4 #&#
The left-hand panel in
Figure~\ref{fig.morediags} plots $\log\operatorname{Var}[W^*]$ against $\log
m$ and
includes a line with the theoretical slope of $-1$ and passing through
an~additional point at $m=1600$. The heavy
left-hand tail at $m=50$ leads to a considerably higher variance than
that which would arise under the SAR; however, even by $m=80$ the fit
is reasonably close.
%f5
%f5 #&#

We assess the degree of dependence of the
distribution of $W^*$ on the position $\mathbf{x}$ by considering the joint
distribution of $W^*$ and $L:=(\log\pi)(\mathbf{X})$, the true
log-target evaluated at
$\mathbf{X}$, where $\mathbf{X}$ is distributed according to the
target. For a
particular $m$, all of
the runs with $\gamma>0$ provide a combined sample of size $n_1$ from the
distribution of the estimate of the log-target at the current
value, $\hat{L}=L+W$, whereas (after scaling so that
$\frac{1}{n_2}\sum_{i=1}^{n_2}\exp{w^{*(i)}}=1$)
each run with $\gamma=0$ provides a sample of size
$n_2$ from the
distribution of
$W^*$ at $\mathbf{x}=\hat{\mathbf{x}}$. Equation \eqref{eqn.joint.stat.w} shows that subject to Assumption~\ref{ass.noise.diff.indep}, $W$ and $L$ are
independent and that the density of $W$ is an exponentially tilted
version of the density of $W^*$. These two properties lead directly to
the following.

%pr3 #&#
\begin{prop}
\label{prop.diagnostic}
If Assumption~\ref{ass.noise.diff.indep} hold, the identity
%
%e4.1 #&#
\begin{equation}
\label{eqn.MGFa}
\mathbb{E} \bigl[ \exp({t\hat{L}}) \bigr] / \mathbb{E} \bigl[
\exp \bigl\{ {(t+1)W^*} \bigr\} \bigr] = \mathbb{E} \bigl[ \exp({tL}) \bigr]
\end{equation}
holds for any $t \in\mathbb{R}$ such that all the above three
expectations are well defined.
\end{prop}

The right-hand side of \eqref{eqn.MGFa} is independent of the noise
distribution, or equivalently of the number of particles,
$m$. Moreover, if the noise is small enough then the ratio on the left-hand
side should provide a good estimator of the true moment generating
function (MGF) of $L$ even if
there is dependence (since the impact of any dependence will be
small).

In our scenario, realisations of $L$ are typically between $-385$ and
$-375$ with a~mode at approximately $-379$, so the MGFs of $L$ and
$\hat{L}$ are
dominated by the term $e^{-379t}$, whatever the noise distribution. To
be able to discern any differences we therefore consider for each
value of $m$, shifted estimators of
the MGFs of $\hat{L}$ and of $L$
\begin{eqnarray*}
M_1(t)&:=& \frac{1}{n_1}\sum_{i=1}^{n_1}
\exp \bigl[t\bigl(\hat{L}^{(i)}+379\bigr) \bigr] \quad\mbox{and}\\
M_2(t) &:=& M_1(t) \Biggl(\frac{1}{n_2}\sum
_{i=1}^n\exp \bigl[(t+1)W^{*(i)} \bigr]
\Biggr)^{-1}.
\end{eqnarray*}
The central panel of Figure~\ref{fig.morediags} shows $M_1(t)$ with a
separate curve for each value of~$m$; the lowest curve is our best estimate of the true MGF of
$L$ ($M_2(t)$ from $m=1600$). The right-hand panel shows $M_2(t)$ for
each value of $m$. Clearly the curves in
the right-hand panel do not coincide, and so the assumption of
independence does not hold precisely. However, it is clear from the
very different vertical scales of the two figures that \textit{most}
of the
difference between the distribution of $\hat{L}$ for any given $m$ and
the distribution of $L$ \textit{can} be explained by Assumption~\ref{ass.noise.diff.indep}.

\begin{figure}

\includegraphics{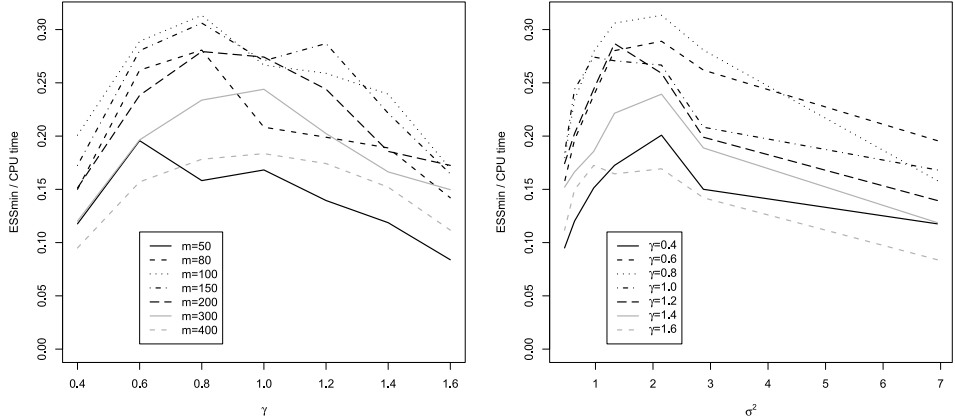}

\caption{Empirical efficiency, $\widehat{\mathrm{eff}}$, measured in terms
of minimum effective sample size per CPU second, plotted against (left
panel) $\gamma$ for different values of $m$ and (right panel)
$\sigma^2$ (estimated from the sample of $W^*$ at the posterior median,
$\hat{\mathbf{x}}$) for different values of $\gamma$.}
\label{fig.effsimstud}
\end{figure}

We now consider an empirical measure of efficiency
$\widehat{\mbox{eff}}$, the quotient of the minimum (over the parameters)
effective sample size and the CPU time. The left-hand panel of Figure~\ref{fig.effsimstud} shows $\widehat{\mbox{eff}}$ plotted against
$\gamma$ for different values of $m$, whilst the right-hand panel
shows $\widehat{\mbox{eff}}$ plotted against
$m$ for different values of $\gamma$. The optimal (over $\mathbb{G}$)
value for $\gamma$ is either $0.8$ or $1.0$ whatever the value of $m$,
which is consistent with the expected insensitivity of the optimal
scaling and suggests that the target is at least approximately
Gaussian. The optimal (over $\mathbb{M}$) value for $m$ is either
$m=200$, $m=150$, or $m=100$, corresponding to an optimal $\sigma^2$ (estimated
from the sample for $W^*$) of either $1.0$, $1.3$ or $2.1$, again (as
far as
can be discerned) showing no strong sensitivity to $\gamma$. Finally
the overall optimum occurs at $\sigma^2=2.1$ and $\gamma=0.8$ with
an acceptance rate of $15.39\%$. The optimal $\sigma^2$ is
slightly lower than the theoretically optimal value of
$3.3$. Further theoretical investigations (using numerical integration)
for a true $5$-dimensional Gaussian target corrupted by noise subject
to the SAR show that ESJD per second is still optimised at $\sigma
^2\approx3.3$; however empirical investigations show that the ESS$/$sec
for this target is optimised at a value of $\sigma^2\approx2$. The
discrepancy between the theory and our simulation study is therefore
likely to be attributable to this discrepancy between ESS and ESJD in
low-dimensional settings. The
relatively high acceptance rate is a consequence of this lower variance
and fits with our theory since from \eqref{eq.closed.form.eff} the
acceptance rate should be $2   \Phi(-\frac{1}{2}\sqrt{2\sigma
^2+\gamma^2 \times2.56^2})=14.7\%$.
%f6
%f6 #&#

\section{Proofs of results}
Equation \eqref{eqn.joint.stat.w} yields that $B=W^*-W$ has density
$\rho$ satisfying
\begin{eqnarray*}
\rho(b) &:=&  \int_{w \in\mathbf{R}} g^*(w) g^*(w+b) e^w \,dw \\
&=&
\int_{w^* \in\mathbf{R}} g^*\bigl(w^*-b\bigr) g^*\bigl(w^*
\bigr)e^{w^*-b} \,dw^* = e^{-b} \rho(-b).
\end{eqnarray*}
Thus
%
%e5.1 #&#
\begin{equation}
\label{eqn.rho.symm}
\qquad\rho(b) = e^{-b/2} h(b)\qquad \mbox{where $h$ is a symmetric
function}, h(b) = h(-b).
\end{equation}
This fact will be used in the proofs of
Theorem~\ref{thm.asymp.analysis} and
Proposition~\ref{prop.func.of.alpha}.

%s5.1 #&#
\subsection{Proof of Theorem~\texorpdfstring{\protect\ref{thm.asymp.analysis}}{1}}\label{sec.proof.main.thm}
For notational convenience, we drop the index $[\cdot]^{(d)}$ when the
context is clear. As in Section~\ref{sect.highdim}, the Hessian matrix
of the log-likelihood $L(\mathbf{x}) := \log\pi^{(d)}(\mathbf{x})$
at $\mathbf{x}\in\mathbf{R}^d$ is denoted by $H(\mathbf{x}) =
[ \partial^2_{ij} L(\mathbf{x}) ]_{1 \leq i,j \leq d}$.
\begin{itemize}
\item \textit{Proof of equation} (\ref{eqn.gen.lim.alpha}).
The mean acceptance probability equals
\begin{eqnarray*}
\alpha^{(d)}(\ell) &:=&  \mathbb{E} \bigl[ 1 \wedge\exp \bigl( L\bigl(
\mathbf{X}+\lambda^{(d)} \mathbf{Z}\bigr)-L(\mathbf{X}) + B \bigr) \bigr]
\\
&=& \mathbb{E} \bigl[ F \bigl( L\bigl(\mathbf{X}+\lambda^{(d)} \mathbf{Z}
\bigr)-L(\mathbf{X}) + B \bigr) \bigr]
\end{eqnarray*}
with $\mathbf{X}\stackrel{\mathcal{D}}{\sim}\pi^{(d)}$, jump scale
$\lambda^{(d)} := \ell/s^{(d)}$,
random variable $\mathbf{Z}\stackrel{\mathcal{D}}{\sim
}\mathbf{N}(0, \mathbf{I}_d)$ independent
from $\mathbf{X}$, and accept-reject function $F(u) := 1 \wedge
\exp(u)$.
Algebra shows that
for any $b \in\mathbf{R}$ and $V \stackrel{\mathcal{D}}{\sim
}\mathbf{N}(-\ell^2/2, \ell^2)$, we have
$\mathbb{E}[1 \wedge\exp(V + b)] = \Phi(-\ell/2 + b/\ell) +
e^{b}\Phi(-\ell/2 - b / \ell)$. By (\ref{eqn.rho.symm})
\begin{eqnarray*}
&& \mathbb{E} \bigl[1\wedge\exp(V+B) \bigr]\\
 && \qquad= \int_{-\infty}^\infty
h(b) \bigl( e^{-b/2}\Phi(-\ell/2 + b/\ell) + e^{b/2}\Phi(-\ell/2
- b/\ell) \bigr) \,db
\\
&&\qquad= 2\int_{-\infty}^\infty h(b) e^{-b/2}\Phi (-
\ell/2 + b/\ell )\,db = 2 \mathbb{E} \bigl[\Phi (-\ell/2 + B/\ell ) \bigr].
\end{eqnarray*}
Since $F$ is continuous and bounded, in order to prove
equation \eqref{eqn.gen.lim.alpha}, it therefore suffices to show that
$L(\mathbf{X}+\lambda^{(d)} \mathbf{Z})-L(\mathbf{X})$ converges in
law to a
Gaussian distribution with mean $-\ell^2 / 2$ and variance $\ell^2$.
A second-order expansion yields
\[
L\bigl(\mathbf{X}+\lambda^{(d)} \mathbf{Z}\bigr)-L(\mathbf{X}) =
\lambda^{(d)} \bigl\langle\nabla L(\mathbf{X}), \mathbf{Z} \bigr\rangle +
\tfrac{1}{2} \bigl( \lambda^{(d)} \bigr)^2 \bigl
\langle\mathbf{Z}, H(\mathbf {X}) \mathbf{Z} \bigr\rangle + R\bigl(\mathbf{X},
\mathbf{Z},\lambda^{(d)}\bigr)
\]
with remainder
$R(\mathbf{X}, \mathbf{Z},\lambda^{(d)}) :=  ( \lambda^{(d)}
 )^2
\int_0^1 (1-t)   \langle\mathbf{Z},   [H( \mathbf{X}+t \lambda
^{(d)} \mathbf{Z})-\break  H(\mathbf{X})]   \mathbf{Z} \rangle   \,dt$.
Slutsky's lemma shows that
to finish the proof of \eqref{eqn.gen.lim.alpha} it suffices to verify that
$\lambda^{(d)} \langle\nabla L(\mathbf{X}), \mathbf{Z} \rangle$
converges in law to a centred Gaussian distribution with variance $\ell
^2$ and that
\[
\lim_{d \to\infty} \tfrac{1}2 \bigl( \lambda^{(d)}
\bigr)^2 \bigl\langle\mathbf{Z}, H(\mathbf{X}) \mathbf{Z} \bigr\rangle
= -\ell^2 / 2 \quad\mbox{and}\quad \lim_{d \to\infty} R\bigl(
\mathbf{X},\mathbf{Z}, \lambda^{(d)}\bigr) = 0
\]
in probability.
\begin{itemize}
\item Note that conditionally upon $\mathbf{X}=\mathbf{x}\in\mathbf
{R}^d$ the quantity $\lambda^{(d)} \langle\nabla L(\mathbf{X}),
\mathbf{Z} \rangle$ has a centred Gaussian distribution with variance
$\ell^2   \|\nabla L(\mathbf{x})\|^2 /  ( s_G^{(d)}  )^2$.
Equation \eqref{eq.rescaled.grad.hess} shows that $\lambda^{(d)}
\langle\nabla L(\mathbf{X}), Z \rangle$ converges in law to a
Gaussian distribution with variance $\ell^2$.
\item Conditionally upon $\mathbf{X}=\mathbf{x}$ the quantity $ (
\lambda^{(d)}  )^2 \langle\mathbf{Z}, H(\mathbf{X})   \mathbf
{Z} \rangle$ has the same distribution as $\ell^2    ( \sum_{i=1}^d \beta_i(\mathbf{x})   Z_i^2  ) / s_L^{(d)}$ where $
( \beta_1(\mathbf{x}), \ldots, \beta_d(\mathbf{x}) )$ is the
spectrum of the Hessian matrix $H(\mathbf{x})$. The conditional mean
thus equals the rescaled Laplacian $\ell^2   {\Delta L(\mathbf
{x})}/{s_L^{(d)}}$, and the conditional variance is
\[
2 \ell^4 \sum_{i=1}^d
\beta_i(\mathbf{x})^2 / \bigl( s_L^{(d)}
\bigr)^2 = 2 \ell^4 \operatorname{Trace} \bigl[
H^2(\mathbf {x}) \bigr] / \bigl( s_L^{(d)}
\bigr)^2.
\]
Markov's inequality, equations \eqref{eq.rescaled.grad.hess} and
\eqref{eqn.eigen.cip}, and the hypothesis $s_L^{(d)} =  (
s_g^{(d)}  )^2$ yield that $\frac{1}2    ( \lambda^{(d)}
)^2 \langle\mathbf{Z}, H(\mathbf{X}) , \mathbf{Z} \rangle$
converges in probability to $-\ell^2/2$.
\item
Equation \eqref{eqn.hessian.regularity} shows that the remainder
$R(\mathbf{X}, \mathbf{Z}, \lambda^{(d)})$ converges to zero in probability.
\end{itemize}

\item \textit{Proof of equation} \eqref{eqn.gen.lim.esjd}.
The proof of equation~\eqref{eqn.gen.lim.esjd}
follows from equation~\eqref{eqn.gen.lim.alpha}.
Note that we have
\begin{eqnarray*}
&& \frac{ (s^{(d)} )^2}{\operatorname{Trace}  [{\mathbf
{T}^{({d})}} ]} \times\mathbb{E} \bigl\| \mathbf{X}^{(d)}_{k+1}
- \mathbf{X}^{(d)}_k \bigr\|_{\mathbf{T}^{(d)}}^2 \\
&& \qquad:=
\ell^2 \mathbb{E} \biggl[ \frac{ \| \mathbf{Z} \|_{\mathbf
{T}^{(d)}}^2}{\operatorname{Trace}  [{\mathbf{T}^{({d})}}
]}\times F \bigl( L\bigl(
\mathbf{X}+\lambda^{(d)} \mathbf{Z}\bigr)-L(\mathbf{X}) + B \bigr)
\biggr].
\end{eqnarray*}
Since $\lim_{d \to\infty} \mathbb{E} [ F ( L(\mathbf
{X}+\lambda^{(d)}
\mathbf{Z})-L(\mathbf{X}) + B )  ] = \alpha(\ell)$, to
prove equation \eqref{eqn.gen.lim.esjd} it suffices to verify that
\[
\mathbb{E} \biggl[ \biggl\{ \frac{ \| \mathbf{Z} \|_{\mathbf
{T}^{(d)}}^2}{\operatorname{Trace}  [{\mathbf{T}^{({d})}}
]} - 1 \biggr\} \times F \bigl( L
\bigl(\mathbf{X}+\lambda^{(d)} \mathbf {Z}\bigr)-L(\mathbf{X}) + B \bigr)
\biggr]
\]
converges to zero as $d \to\infty$.
Since the function $F$ is bounded, the conclusion follows once we have
proved that
$\mathbb{E} [ ( {\| \mathbf{Z}\|_{\mathbf
{T}^{(d)}}^2}/{\operatorname{Trace}  [{\mathbf{T}^{({d})}}
]} - 1  )^2 ]$ converges to zero. Diagonalisation of the
symmetric matrix $\mathbf{T}^{(d)}$ in an orthonormal basis shows that
this last quantity equals
$2 \times\operatorname{Trace}  [{ ( \mathbf{T}^{({d})}
)^2} ] /\break 
\operatorname{Trace}  [{\mathbf{T}^{({d})}} ]^2$ so that the conclusion directly follows from equation
\eqref{eqn.eccentric.distance}.
\end{itemize}

%s5.2 #&#
\subsection{Proof of Proposition~\texorpdfstring{\protect\ref{prop.func.of.alpha}}{1}}
\label{sec.proof.bijection.}

The dominated convergence theorem shows that $\ell\mapsto\alpha(\ell
) = 2 \times\mathbb{E} [ \Phi(B/\ell- \ell/2)  ]$ is
continuous and converges to zero as $\ell$ tends to infinity. Since
the limiting acceptance probability can also be expressed as $\alpha
(\ell) = 2   \mathbb{P}(\ell  \xi+ \ell^2/2 < B)$ for $\xi
\stackrel{\mathcal{D}}{\sim}\mathbf{N}(0,1)$ independent from all
other sources of randomness, it also follows that the limiting
acceptance probability $\alpha(\ell)$ converges to $2   \mathbb{P}(B>0)$
as $\ell$ converges to zero. To finish the proof of
Proposition~\ref{prop.func.of.alpha}, it remains to verify that the
function $\ell\to\alpha(\ell)$ is strictly decreasing. To this end,
we will establish that the derivative $\frac{d}{d\ell} \alpha(\ell
)$ is strictly negative.
Applying (\ref{eqn.rho.symm}), the derivative of $\ell\mapsto\alpha
(\ell)$ is
\begin{eqnarray*}
\frac{d\alpha}{d\ell}(\ell) &=&  \frac{d}{d\ell} \int_{b \in\mathbf{R}} 2
\Phi [ -\ell/ 2 + b/ \ell ] e^{-b/2} h(b) \,db \\
&=&  -\int_{b \in\mathbf{R}}
\varphi [ -\ell/ 2 + b/ \ell ] \biggl\{1 + \frac{2b}{\ell^2} \biggr\}
e^{-b/2} h(b) \,db
\end{eqnarray*}
with $\varphi(x)=\Phi'(x) = e^{-x^2/2} / \sqrt{2 \pi}$ the density
of a standard Gaussian distribution. Algebra shows that the function $b
\mapsto b   e^{-b/2}   \varphi[ -\ell/ 2 + b/ \ell]$ is odd so
that the derivative simplifies,
\[
\frac{d\alpha}{d\ell}(\ell) = - \int_{b \in\mathbf{R}} \varphi [ -\ell/ 2 +
b/ \ell ] e^{-b/2} h(b) \,db.
\]
This quantity is clearly strictly negative, completing the proof
of Proposition~\ref{prop.func.of.alpha}.

%
%s5.3 #&#
\subsection{Proof of Proposition~\texorpdfstring{\protect\ref{prop.rel.eff.bds}}{2}}
\label{proof.prop.bds}
The upper bound follows from a similar argument to that in
\cite{AndrieuVihola2014}. Let $\widetilde{W}$ be an independent copy
of $W^*$, and let $V \stackrel{\mathcal{D}}{\sim}\mathbf{N}(-\ell
^2/2, \ell^2)$ be independent from any other source of randomness.
Relating $\widetilde{W}$ to $W$ through \eqref{eqn.joint.stat.w} yields
\begin{eqnarray*}
\mathbb{E} \bigl[{1\wedge\exp(V+B)} \bigr] %=\Expect{1\wedge\exp(V+W^*-W)}
&=& \mathbb{E} \bigl[{\exp(
\widetilde{W})\wedge \exp(V)\exp \bigl(W^*\bigr)} \bigr]\\
&\le & \mathbb{E} \bigl[{1
\wedge\exp(V)} \bigr]=2 \times\Phi(-\ell/2);
\end{eqnarray*}
we have applied Jensen's inequality twice
to the function $(x,y) \mapsto x\wedge\exp(V)y$ which is concave in
both $x$ and $y$. Since $J(\ell) = \mathbb{E}  [{1\wedge\exp
(V+B)} ]$, the upper bound follows.

The lower bound follows from a similar argument to that used in
\cite{Doucetetal2014}.
We note that $(1\wedge e^V)(1\wedge e^B)\le
1\wedge e^{V+B}$. $V$ and $B$ are independent by assumption; as $\alpha_{\max} = \mathbb{E}[1\wedge e^B]$, the
result follows on taking expectations with respect to both of these variables.

%
%s5.4 #&#
\subsection{Proof of Theorem \texorpdfstring{\protect\ref{thm.diff.lim}}{2}} \label{sec.proof.diff.lim}
In this section we use the following notation.
We write $u_n \lesssim v_n$ when the absolute value
of the quotient $u_n / v_n$ is bounded above by a constant which is
independent of the index $n$; we write\vspace*{1pt} $u_n \asymp v_n$ if $u \lesssim
v_n$ and $v_n \lesssim u_n$.
For $(\mathbf{x},w) \in\mathbf{R}^d \times\mathbf{R}$ we write
$\mathbb{E}_{\mathbf{x}, w}[  \cdot ]$
instead of $\mathbb{E}[   \cdot  | (\mathbf
{X}^{(d)}_0,W^{(d)}_0)=(\mathbf{x},w)]$.
The Metropolis--Hastings accept-reject function is the globally
Lipschitz function $F(u) = 1 \wedge e^u$. The log-likelihood function
is denoted by $A := \log f$ in this section. We drop the index $(\cdot
)^{(d)}$ when the context is clear.

The proof follows ideas from \cite{BedardA2007}, which itself is an
adaptation of the original paper \cite{RobertsGelmanGilks1997}.
It is based on \cite{ethier1986markov}, Theorem $8.2$, Chapter $4$,
which gives conditions under which
the finite dimensional distributions of a sequence of
processes converge weakly to those of some Markov process.
\cite{ethier1986markov}, Corollary $8.6$, Chapter $8$,
provides further conditions for this sequence
of processes to be relatively compact in the appropriate topology
and thus establish weak convergence of the
stochastic processes themselves.

The situation is slightly more involved than the one presented in
\cite{RobertsGelmanGilks1997,BedardA2007};
the proof needs a homogenisation argument since the processes
$\mathbf{X}^{(d)}$ and $W^{(d)}$ evolve on two different time scales.
Indeed, it will become apparent from the proof that the process
$\mathbf{X}^{(d)}$ takes $\mathcal{O}(d)$ steps to mix while the
process $W^{(d)}$ takes
$\mathcal{O}(1)$ steps to mix. In order to exploit this time-scales
separation, we introduce an intermediary time scale $T_d = \lfloor
{d^{\gamma}} \rfloor$
where $0<\gamma<1/4$ is an exponent whose exact value is not important
to the proof. The intuition is that after $\mathcal{O}(T_d)$ steps
the process $W^{(d)}$ has mixed while each coordinate of $\mathbf
{X}^{(d)}$ has only moved by an infinitesimal quantity.
We introduce the subsampled processes
$\widetilde{\mathbf{X}}^{(d)}$ and $\widetilde{W}^{(d)}$ defined by
\[
\widetilde{\mathbf{X}}^{(d)}_{k} = \mathbf{X}^{(d)}_{kT_d}
\quad\mbox{and}\quad \widetilde{W}^{(d)}_k = W^{(d)}_{k T_d}.
\]
One\vspace*{1pt} step of the process $\widetilde{\mathbf{X}}^{(d)}$
(resp.,
$\widetilde{W}^{(d)}$) corresponds to $T_d$ steps of the process
$\mathbf{X}^{(d)}$ (resp., $W^{(d)}$).
We then define an accelerated version $\widetilde{V}^{(d)}$ of the
subsampled first coordinate\vspace*{-1.5pt} process $k \mapsto\widetilde
{X}^{(d)}_{k,1}$. In order to prove a diffusion limit
for the first coordinate of the process $\mathbf{X}^{(d)}$, one needs
to accelerate time by a factor of $d$; consequently, in order to prove
a diffusion limit for the process
$\widetilde{\mathbf{X}}^{(d)}$, one needs to accelerate time by a
factor $d / T_d$, and thus define $\widetilde{V}^{(d)}$ by
\[
\widetilde{V}^{(d)}(t) := \widetilde{X}^{(d)}_{\lfloor{td/T_d}
\rfloor,1}.
\]
The proof then consists of showing that the sequence $\widetilde
{V}^{(d)}$ converges weakly in the Skorohod topology towards the
limiting diffusion \eqref{e.limiting.diffusion} and verifying that $\|
\widetilde{V}^{(d)} - V^{(d)}\|_{\infty, [0,T]}$ converges\vspace*{1pt} to zero in
probability;
this is enough to prove that the sequence $V^{(d)}$ converges weakly in
the Skorohod topology towards the
limiting diffusion \eqref{e.limiting.diffusion}. The proof is divided
into three main steps. First, we show
that the finite dimensional marginals of the process $\widetilde
{V}^{(d)}$ converge to those of the limiting diffusion \eqref{e.limiting.diffusion}.
Second, we establish that the sequence $\widetilde{V}^{(d)}$ is weakly
relatively compact. These two steps
prove that the sequence $\widetilde{V}^{(d)}$ converges weakly in the
Skorohod topology\vspace*{1.5pt} towards the diffusion \eqref{e.limiting.diffusion}.
As a final step, we prove that the quantity $\| \widetilde{V}^{(d)} -
V^{(d)}\|_{\infty, [0,T]}$ converges to zero in probability, establishing
the weak convergence of the sequence $V^{(d)}$ towards the diffusion
\eqref{e.limiting.diffusion}.
Before embarking on the
proof we define several quantities that will be needed in the sequel.
We denote\vspace*{1.5pt} by $\mathcal{L}$ the generator
of the limiting diffusion $\eqref{e.limiting.diffusion}$. Similarly,
we define $\mathcal{L}^{(d)}$ and $\widetilde{\mathcal{L}}^{(d)}$
the approximate generators\vspace*{1pt}
of the first coordinate process $\{X^{(d)}_{k,1} \}_{k \geq0}$ and its
accelerated version $\{\widetilde{X}^{(d)}_{k,1} \}_{k \geq0}$; for
any smooth and compactly supported test function $\varphi\dvtx \mathbf
{R}\to\mathbf{R}$,
vector $\mathbf{x}=(x_1, \ldots,x_d) \in\mathbf{R}^d$ and scalar $w
\in\mathbf{R}$, we have
\[
\cases{\ds\mathcal{L}\varphi(x_1) =
\tfrac{1}{2} h(\ell) \bigl[ \varphi^{\prime\prime}(x_1) +
A(x_1) \varphi'(x_1) \bigr],
\vspace*{5pt}\cr
\ds \mathcal{L}^{(d)} \varphi(\mathbf{x},w) = \mathbb{E}_{\mathbf{X}^{(d)},W}
\bigl[ \varphi\bigl(X^{(d)}_{1,1}\bigr) - \varphi(x_1)
\bigr] / \delta,
\vspace*{5pt}\cr
\ds \widetilde{\mathcal{L}}^{(d)} \varphi(\mathbf{x},w) =
\mathbb{E}_{\mathbf{X}^{(d)},W} \bigl[ \varphi\bigl(\widetilde{X}^{(d)}_{1,1}
\bigr) - \varphi(x_1) \bigr] / (T_d \times\delta )}
\]
with $\delta= 1/d$.
Note that although $\varphi$ is a scalar function, the functions
$\mathcal{L}^{(d)} \varphi$ and $\widetilde{\mathcal
{L}}^{(d)}\varphi$ are defined on $\mathbf{R}^d \times\mathbf{R}$.
In the sequel we sometimes write $\widetilde{\mathcal{L}}^{(d)}
\varphi(x_1, \ldots, x_d,w)$ instead of $\widetilde{\mathcal
{L}}^{(d)} \varphi(\mathbf{x},w)$.

%
%s5.4.1 #&#
\subsubsection{Convergence of the finite dimensional distributions of
\texorpdfstring{$\widetilde{V}^{(d)}$}{widetilde{V}{(d)}}}
In this section we prove that the finite dimensional distributions of
the sequence of processes $\widetilde{V}^{(d)}$ converge weakly to
those of the diffusion \eqref{e.limiting.diffusion}.
Since the limiting process is a scalar diffusion, the set of smooth and
compactly supported functions
is a core for the generator of the limiting diffusion (\cite{ethier1986markov},
Theorem $2.1$, Chapter $8$); in the sequel, one can
thus work with test functions belonging to this core only.
To prove the convergence of the finite dimensional marginals, one can
apply \cite{ethier1986markov}, Chapter~$4$, Theorem $8.2$, Corollary~$8.4$, to the pair $(\xi^{(d)},\varphi^{(d)})$ defined by
%
%e5.2 #&#
\begin{eqnarray}
\xi^{(d)}(t) &=&  \frac{1}{\delta  T_d} \int
_{t}^{t + \delta
T_d} \varphi\bigl[\widetilde{V}^{(d)}(s)
\bigr] \,ds\quad \mbox{and}
\nonumber
\\[-8pt]
\label{eq.xi.phi}\\[-8pt]
\nonumber
\varphi^{(d)}(t) &=&  \widetilde{\mathcal{L}}^{(d)}
\varphi \bigl( \widetilde{\mathbf{X}}^{(d)}_{\lfloor{td/T_d} \rfloor} , \widetilde
{W}^{(d)}_{\lfloor{td/T_d} \rfloor} \bigr).
\end{eqnarray}
To establish that this result applies, we will concentrate on proving
that for any smooth and compactly supported function $\varphi\dvtx \mathbf
{R}\to\mathbf{R}$ the following limit holds:
%
%e5.3 #&#
\begin{equation}
\label{e.finite.dim.conv}
\lim_{d \to\infty} \mathbb{E} \bigl| \widetilde{
\mathcal{L}}^{(d)} \varphi(X_1, \ldots, X_d,W) -
\mathcal{L}\varphi(X_1) \bigr| = 0,
\end{equation}
for $\{X_k\}_{k \geq1}$ an i.i.d. sequence of random variables
distributed according to $f(x) \,dx$
and $W \stackrel{\mathcal{D}}{\sim}e^w g^*(w)   \,dw$. Equation \eqref{e.finite.dim.conv} implies equation (8.11) of
\cite{ethier1986markov}, Chapter~4, and the stationarity assumption
implies equations (8.8) and (8.9) of
\cite{ethier1986markov}, Chapter~4.
To verify that equation (8.10) of
\cite{ethier1986markov}, Chapter~4, holds, one can notice that for
any index $k \geq1$ we have $\mathbb{E} [  | \varphi
(X^{(d)}_{k,1}) - \varphi(X^{(d)}_{0,1})  |  ] \lesssim k
\delta^{1/2}$, which is a direct consequence of the triangle
inequality and the fact that $\varphi$ is a Lipschitz function.
The proof of \eqref{e.finite.dim.conv} is based on an averaging
argument\vspace*{1pt} that exploits the following relationship between
the generators $\mathcal{L}^{(d)}$ and $\widetilde{\mathcal{L}}^{(d)}$,
%
%e5.4 #&#
\begin{equation}
\label{e.telescopin}
\widetilde{\mathcal{L}}^{(d)} \varphi(\mathbf{x},w) =
\mathbb {E}_{\mathbf{x},w} \Biggl[ \frac{1}{T_d} \sum
_{k=0}^{T_d-1} \mathcal{L}^{(d)} \varphi\bigl(
\mathbf{X}^{(d)}_k, W^{(d)}_k\bigr)
\Biggr].
\end{equation}
Equation~\eqref{e.telescopin} follows from the telescoping expansion
$\varphi(\mathbf{X}^{(d)}_{T_d})-\varphi(\mathbf{X}^{(d)}_0) = \sum_{k=0}^{T_d-1} \varphi(\mathbf{X}^{(d)}_{k+1}) - \varphi(\mathbf
{X}^{(d)}_k)$
and the law of iterated conditional expectations.
The following lemma is crucial:
%
% AVERAGING LEMMA

%le1 #&#
\begin{lemma}[(Asymptotic expansion of
$\mathcal{L}^{(d)} \varphi$)]\label{lem.asymp.gen}
Let Assumptions \ref{ass.noise.diff.indep} and \ref{assump.f} be satisfied.
There exist two bounded and continuous functions $a,b\dvtx \mathbf{R}\to
\mathbf{R}$ satisfying the following properties:
\begin{longlist}[(2)]
\item[(1)]
Let\vspace*{-1.5pt} $W$ be a random variable distributed as the stationary distribution
of the log-noise, $W \stackrel{\mathcal{D}}{\sim}e^w   g^*(w)
\,dw$, and $\alpha(\ell)$ be the asymptotic mean acceptance probability
identified in Theorem~\ref{thm.asymp.analysis}. The following identity holds:
%
%e5.5 #&#
\begin{equation}
\label{a.averaging.identity}
\mathbb{E}\bigl[a(W)\bigr] = \mathbb{E}\bigl[b(W)\bigr]=
\tfrac{1}{2} \alpha(\ell).
\end{equation}
\item[(2)]
For any\vspace*{1pt} smooth and compactly supported
function $\varphi\dvtx \mathbf{R}\to\mathbf{R}$ the averaged generator
$\mathcal{G}\varphi$ defined for any $(x_1,w) \in\mathbf{R}^2$ by
\[
\mathcal{G}\varphi(x_1,w) := \frac{\ell^2}{I} \bigl[ a(w)
A'(x_1) \varphi'(x_1) + b(w)
\varphi^{\prime\prime}(x_1) \bigr]
\]
satisfies
\[
\lim_{d \to\infty} \mathbb{E} \bigl| \mathcal{L}^{(d)}
\varphi(X_1, \ldots,X_d,W) - \mathcal{G}
\varphi(X_1,W) \bigr|^2 = 0
\]
for an i.i.d. sequence $\{X_k\}_{k \geq1}$ marginally distributed as
$f(x) \,dx$ and constant $I$ defined by \eqref{e.I}.
\end{longlist}
\end{lemma}

$\!$The above lemma thus shows that the approximate generator $\mathbb
{E}_{\mathbf{X}^{(d)},W} \times\break [ \varphi(X^{(d)}_{1,1}) - \varphi
(x_1) ] / \delta$ asymptotically only depends on the first
coordinate $x_1 \in\mathbf{R}$ and the log-noise $w \in\mathbf{R}$.
The proof is an averaging argument for the $(d-1)$ coordinates $(x_2,
\ldots, x_d)$;
this is mainly technical and details can be
found in Appendix~\ref{sec.asymp.gen.lem}.
The next step consists in exploiting the separation of scales between
the processes $\{ \mathbf{X}^{(d)}_k\}_{k \geq0}$
and $\{W^{(d)}_k\}_{k \geq0}$.

%le2 #&#
\begin{lemma} \label{lem.separation.scale}
Let $h\dvtx \mathbf{R}\to\mathbf{R}$ be a bounded measurable function.
Suppose that for any $d \geq1$ the Markov chain $\{(\mathbf
{X}^{(d)}_{k}, W^{(d)}_{k})\}_{k \geq0}$ is started at stationarity.
The following limit holds:
\[
\lim_{d \to\infty} \mathbb{E} \Biggl| \frac{1}{T_d}\sum
_{k=0}^{T_d-1} h\bigl(W^{(d)}_{k}
\bigr) - \mathbb{E}\bigl[h(W)\bigr] \Biggr| = 0,
\]
with $W$ distributed according to the stationary distribution
$W \stackrel{\mathcal{D}}{\sim}e^w g^*(w)   \,dw$.
\end{lemma}

The above lemma thus shows that $T_d = \lfloor{d^{\gamma}} \rfloor$
steps, with $0<\gamma<1/4$, are enough for the process $W^{(d)}$ to mix.
The proof relies on a coupling argument and the ergodic theorem for
Markov chains.
Details can be found Appendix~\ref{sec.sep.sc.lem}. We now have all
the tools in hands
to prove equation \eqref{e.finite.dim.conv}.
First, with the notation $\mathbf{X}^{(d)}=(X_1, \ldots,X_d)$,
the telescoping expansion \eqref{e.telescopin}
and Jensen's conditional inequality yields
\begin{eqnarray*}
&& \mathbb{E} \bigl| \widetilde{\mathcal{L}}^{(d)} \varphi\bigl(\mathbf
{X}^{(d)},W\bigr) - \mathcal{L}\varphi(X_1) \bigr|\\
&&\qquad = \mathbb{E}
\Biggl| \mathbb{E}_{\mathbf{X}^{(d)},W} \Biggl[ \frac
{1}{T_d} \sum
_{k=0}^{T_d-1} \mathcal{L}^{(d)} \varphi\bigl(
\mathbf {X}^{(d)}_{k}, W^{(d)}_{k}\bigr)
- \mathcal{L}\varphi\bigl(X^{(d)}_{0,1}\bigr) \Biggr] \Biggr|
\\
&&\qquad\leq  \mathbb{E} \Biggl| \frac{1}{T_d} \sum_{k=0}^{T_d-1}
\mathcal {L}^{(d)} \varphi\bigl(\mathbf{X}^{(d)}_{k},
W^{(d)}_{k}\bigr) - \mathcal {L}\varphi\bigl(X^{(d)}_{0,1}
\bigr) \Biggr|.
\end{eqnarray*}
One can then use the triangle inequality to obtain the bound
\begin{eqnarray*}
\label{e.triangle.ineq.gen}
&& \mathbb{E} \bigl| \widetilde{\mathcal{L}}^{(d)} \varphi\bigl(
\mathbf {X}^{(d)},W\bigr) - \mathcal{L}\varphi(X_1) \bigr| \\
&&\qquad\leq
\mathbb{E} \Biggl| \frac{1}{T_d} \sum_{k=0}^{T_d-1}
\mathcal {L}^{(d)} \varphi\bigl(\mathbf{X}^{(d)}_k,
W^{(d)}_{k}\bigr) - \mathcal {L}\varphi\bigl(X^{(d)}_{0,1}
\bigr) \Biggr|
\\
&&\qquad\leq \frac{1}{T_d} \sum_{k=0}^{T_d-1}
\mathbb{E} \bigl| \mathcal {L}^{(d)} \varphi\bigl(\mathbf{X}^{(d)}_k,
W^{(d)}_{k}\bigr) - \mathcal {G}\varphi\bigl(X^{(d)}_{k,1},
W^{(d)}_{k}\bigr) \bigr|
\\
&&\qquad\quad{}+ \mathbb{E} \Biggl| \frac{1}{T_d} \sum_{k=0}^{T_d-1}
\mathcal {G}\varphi\bigl(X^{(d)}_{k,1}, W^{(d)}_{k}
\bigr) - \mathcal{G}\varphi \bigl(X^{(d)}_{0,1},
W^{(d)}_{k}\bigr) \Biggr|
\\
&&\qquad\quad{}+ \mathbb{E} \Biggl| \frac{1}{T_d} \sum_{k=0}^{T_d-1}
\mathcal {G}\varphi\bigl(X^{(d)}_{0,1}, W^{(d)}_{k}
\bigr) - \mathcal{L}\varphi \bigl(X^{(d)}_{0,1}\bigr) \Biggr|
\\
&&\qquad=: E_1(d) + E_2(d) + E_3(d).
\end{eqnarray*}
To complete the proof of the convergence of the finite dimensional
distributions of $\widetilde{V}^{(d)}$
towards those of the limiting diffusion \eqref{e.limiting.diffusion},
it remains to prove that $E_i(d) \to0$ as $d\to\infty$ for $i=1,2,3$:
\begin{itemize}
\item
Since the Markov chain $\{ (\mathbf{X}^{(d)}_{k},W^{k,}) \}_{k \geq
0}$ is assumed to be stationary,
the quantity $E_1(d)$ also equals $\mathbb{E} |   \mathcal
{L}^{(d)} \varphi(X_1, \ldots, X_d, W) - \mathcal{G}\varphi(X_1,
W)   |$.
Lemma~\ref{lem.asymp.gen} shows that $E_1(d) \to0$ as $d \to\infty$.
\item The formula for the quantity $\mathcal{G}\varphi(x,w)$ shows
that the expectation $E_2(d)$ also reads
%
%e5.6 #&#
\begin{eqnarray}
&& \frac{\ell^2}{I} \times\mathbb{E} \Biggl| \frac{1}{T_d} \sum
_{k=0}^{T_d-1} a\bigl(W^{(d)}_{k}
\bigr) \bigl\{ A'\bigl(X^{(d)}_{k,1}\bigr) \varphi
'\bigl(X^{(d)}_{k,1}\bigr) - A'
\bigl(X^{(d)}_{0,1}\bigr) \varphi'
\bigl(X^{(d)}_{0,1}\bigr) \bigr\}
\nonumber
\\[-8pt]
\label{eq.E2.expanded}\\[-8pt]
\nonumber
&&\hspace*{64pt}\quad\qquad{}+ \frac{1}{T_d} \sum_{k=0}^{T_d-1} b
\bigl(W^{(d)}_{k}\bigr) \bigl\{ \varphi ^{\prime\prime}
\bigl(X^{(d)}_{k,1}\bigr) - \varphi^{\prime\prime}
\bigl(X^{(d)}_{0,1}\bigr) \bigr\} \Biggr|.
\end{eqnarray}
Under Assumption \ref{assump.f} the function $A'$ is globally
Lipschitz; since $\varphi$ is smooth with compact support,
the functions $x \mapsto A'(x) \varphi'(x)$ and $x \mapsto\varphi
^{\prime\prime}$ are both globally Lipschitz. Using the boundedness of the functions
$a$ and $b$, this yields that the quantity\vspace*{1pt} in equation \eqref{eq.E2.expanded} is bounded by a constant multiple of
$\frac{1}{T_d} \sum_{k=0}^{T_d-1} \mathbb{E} | X^{(d)}_{k,1} -
X^{(d)}_{0,1}  |$.
For any index $k \geq0$ we have $\mathbb{E} | X^{(d)}_{k+1,1} -
X^{(d)}_{k,1} | \lesssim\delta^{1/2}$
so that $\mathbb{E} | X^{(d)}_{k,1} - X^{(d)}_{0,1}  |
\lesssim k   \delta^{1/2}$. Since $T_d / d^{1/2} \to0$, the
conclusion follows.
\item
Lemma~\ref{lem.asymp.gen} shows that one can express
the generator of the limiting diffusion~\eqref{e.limiting.diffusion} as
$\mathcal{L}\varphi(x) = \frac{\ell^2}{I} \mathbb{E}[a(W)]
A'(x)   \varphi'(x) + \frac{\ell^2}{I} \mathbb{E}[b(W)]   \varphi
^{\prime\prime}(x)$.
The expectation $E_3(d)$ thus also reads
\begin{eqnarray*}
&& \frac{\ell^2}{I} \times  \mathbb{E} \Biggl| \Biggl\{\frac{1}{T_d} \sum
_{k=0}^{T_d-1} a\bigl(W^{(d)}_{k}
\bigr) - \mathbb{E}\bigl[a(W)\bigr] \Biggr\} A'\bigl(X^{(d)}_{0,1}
\bigr) \varphi'\bigl(X^{(d)}_{0,1}\bigr)
\\
&&\hspace*{28pt}\qquad\quad{}+ \Biggl\{ \frac{1}{T_d} \sum_{k=0}^{T_d-1}
b\bigl(W^{(d)}_{k}\bigr) - \mathbb {E}\bigl[b(W)\bigr]
\Biggr\} \varphi^{\prime\prime}\bigl(X^{(d)}_{0,1}\bigr) \Biggr|.
\end{eqnarray*}
Because the function $\varphi$ is smooth with compact support, it
follows\break (Cauchy--Schwarz) that this quantity is less than a constant
multiple of
\begin{eqnarray*}
&& \mathbb{E} \Biggl[ \Biggl\{ \frac{1}{T_d} \sum
_{k=0}^{T_d-1} a\bigl(W^{(d)}_{k}
\bigr) - \mathbb{E}\bigl[a(W)\bigr] \Biggr\}^2 \Biggr]^{1/2}
\times \mathbb{E} \bigl[ A'(X)^2 \bigr]^{1/2} \\
&& \qquad{}+
\mathbb{E} \Biggl| \frac{1}{T_d} \sum_{k=0}^{T_d-1}
b\bigl(W^{(d)}_{k}\bigr) - \mathbb{E}\bigl[b(W)\bigr] \Biggr|.
\end{eqnarray*}
Lemma~\ref{lem.separation.scale} shows that $\mathbb{E} | \frac
{1}{T_d} \sum_{k=0}^{T_d-1} b(W^{(d)}_{k}) - \mathbb{E}[b(W)]  |
\to0$,
and under Assumption \ref{assump.f} the expectation $\mathbb{E}
[ A'(X)^2 ]$ is finite. Therefore, to finish\break the proof of the limit
$E_3(d) \to0$, one needs to verify that\break $\mathbb{E} [  \{
\frac{1}{T_d} \sum_{k=0}^{T_d-1} a(W^{(d)}_{k}) - \mathbb{E}[a(W)]
 \}^2 ] \to0$.
According to Lemma~\ref{lem.separation.scale}, the sequence $
(\frac{1}{T_d} \sum_{k=0}^{T_d-1} a(W^{(d)}_{k}) - \mathbb{E}[a(W)]
 )$
converges in $L^1$ to zero. The sequence is also bounded in $L^{\infty
}$ since the function $a$ is bounded. A sequence bounded in $L^{\infty
}$ that converges to zero in~$L^1$ also converges to zero in any $L^p$
for $1 \leq p < \infty$. The conclusion follows.
\end{itemize}

%s5.4.2 #&#
\subsubsection{Relative weak compactness of the sequence \texorpdfstring{$\widetilde{V}^{(d)}$}{widetilde{V}{(d)}}}
The process $\widetilde{V}^{(d)}$ is started at stationarity
and the space of smooth functions with compact support is an
algebra that strongly separates points.
Ethier and Kurtz (\cite{ethier1986markov}, Chapter~4,
Corollary~8.6) show that
in order to prove that the sequence $\widetilde{V}^{(d)}$ is
relatively weakly compact in the
Skorohod topology it suffices to verify that equations (8.33) and
(8.34) of \cite{ethier1986markov}, Chapter~4, hold.
\begin{itemize}
\item To prove (8.34) it suffices to show that for any smooth and
compactly supported test function $\varphi$ the sequence $d \mapsto
\mathbb{E} | \widetilde{\mathcal{L}}^{(d)} \varphi(X_1, \ldots
, X_d,W)  |^2$ is bounded. One can use the telescoping\vspace*{1pt} expansion
\eqref{e.telescopin}, Lemma~\ref{lem.asymp.gen}
and the stationarity of the Markov chain $\{(\mathbf{X}^{(d)}_{k},
W^{(d)}_{k})\}_{k \geq0}$
and obtain that
\begin{eqnarray*}
\mathbb{E} \bigl| \widetilde{\mathcal{L}}^{(d)} \varphi\bigl(\mathbf
{X}^{(d)},W\bigr) \bigr|^2  &\lesssim &  \mathbb{E} \Biggl|
\frac{1}{T_d}\sum_{k=0}^{T_d-1}
\mathcal{L}^{(d)} \varphi\bigl(\mathbf{X}^{(d)}_k,W^{(d)}_{k}
\bigr) - \mathcal{G}\varphi \bigl(X^{(d)}_{k,1},
W^{(d)}_{k}\bigr) \Biggr|^2
\\[2pt]
&&{}+ \mathbb{E} \Biggl| \frac{1}{T_d}\sum_{k=0}^{T_d-1}
\mathcal{G}\varphi \bigl(X^{(d)}_{k,1}, W^{(d)}_{k}
\bigr) \Biggr|^2
\\[2pt]
&\leq & \frac{1}{T_d}\sum_{k=0}^{T_d-1}
\mathbb{E} \bigl| \mathcal{L}^{(d)} \varphi\bigl(\mathbf{X}^{(d)},W^{(d)}_{k}
\bigr) - \mathcal{G}\varphi \bigl(X^{(d)}_{k,1},
W^{(d)}_{k}\bigr) \bigr|^2 \\[2pt]
&&{}+ \frac{1}{T_d}\sum
_{k=0}^{T_d-1} \mathbb{E} \bigl| \mathcal{G}\varphi
\bigl(X^{(d)}_{k,1}, W^{(d)}_{k}\bigr)
\bigr|^2
\\[2pt]
&=& \mathbb{E} \bigl| \mathcal{L}^{(d)} \varphi\bigl(\mathbf{X}^{(d)},W
\bigr) - \mathcal{G}\varphi(X_1, W) \bigr|^2 + \mathbb{E} \bigl|
\mathcal {G}\varphi(X_1, W) \bigr|^2 \\
&=& o(1) + \mathcal{O}(1).
\end{eqnarray*}
This proves equation~(8.34).
\item To\vspace*{1pt} prove (8.33) one needs to show that the expectation of $\sup
 \{ | \xi_d(t) - \widetilde{V}^{(d)}(t) |  \dvtx    t
\in[0,T] \}$ converges to zero as $d \to\infty$, where the
process $\xi_d$ is defined in equation \eqref{eq.xi.phi}. Note that
the supremum is less than
%
%e5.7 #&#
\begin{equation}
\label{eq.discrepancy}
\quad\| \varphi\|_{\mathrm{Lip}} \times\sup \Biggl\{ \delta\times\sum
_{k=i}^j \bigl| X^{(d)}_{j,1}
- X^{(d)}_{i,1} \bigr| \dvtx  0 \leq i < j \leq d \times T \mbox{ and }
|i-j| \leq T_d \Biggr\},\hspace*{-10pt}
\end{equation}
where $\| \varphi\|_{\mathrm{Lip}}$ is the Lipschitz constant of
$\varphi$. Therefore, since $ | X^{(d)}_{j,1} - X^{(d)}_{i,1}
| \lesssim\delta  \sum_{k=i}^{j-1} |Z_{k}|$ where $\{Z_{k}\}_{k
\geq0}$ are i.i.d. standard Gaussian random variables such that
$X^{(d),*}_{i,1}=X^{(d)}_{i,1} + \ell  I^{-1/2}   \delta  Z_{k}$,
the following lemma gives the conclusion.
\end{itemize}

%le3 #&#
\begin{lemma} \label{lem.discrepancy}
Let $\{\xi_k\}_{k \geq1}$ an i.i.d. sequence of standard Gaussian
random variables $\mathbf{N}(0,1)$. We have
\[
\lim_{d \to\infty} \mathbb{E} \Biggl[ \sup \Biggl\{ \delta\times\sum
_{k=i}^j | \xi_k | \dvtx  0 \leq
i < j \leq d \times T \mbox{ and } |i-j| \leq T_d \Biggr\} \Biggr] = 0.
\]
\end{lemma}

\begin{pf} Indeed, it suffices to prove that $\delta$ times the
expectation of the supremum $\sup\{ S(i,d) \dvtx  i \leq d / T_d\}$, with
$S(i,d) = \sum_{k=i   T_d}^{(i+1)   T_d}  | \xi_k  |$,
converges to zero; this follows from Markov's inequality and standard
Gaussian computations.
\end{pf}

This completes the proof of the relative weak compactness in the
Skorohod topology. The sequence of processes $\widetilde{V}^{(d)}$ is
weakly compact in the Skorohod topology, and the finite dimensional
distributions of $\widetilde{V}^{(d)}$ converge to the finite
dimensional distribution of the diffusion \eqref
{e.limiting.diffusion}. Consequently, the sequence of processes
$\widetilde{V}^{(d)}$ converges weakly in the Skorohod space
$D([0,T])$ to the diffusion~\eqref{e.limiting.diffusion}. The next
section shows that the discrepancy between $V^{(d)}$ and $\widetilde
{V}^{(d)}$ is small and thus proves that the sequence of processes
$V^{(d)}$ also converges to the diffusion \eqref{e.limiting.diffusion}.
%
%s5.4.3 #&#
\subsubsection{Discrepancy between \texorpdfstring{$V^{(d)}$}{V{(d)}} and \texorpdfstring{$\widetilde{V}^{(d)}$}{widetilde{V}{(d)}}}
Since $\sup_{t \leq T}    | V^{(d)}_t - \widetilde{V}^{(d)}_t
 |$ is less than the supremum of equation \eqref{eq.discrepancy},
Lemma~\ref{lem.discrepancy} yields that $\|\widetilde{V} -\break  V^{(d)}\|
_{\infty,[0,T]}$ converges to zero in probability.
This ends the proof of Theorem~\ref{thm.diff.lim}.

%
%s6 #&#
\section{Discussion}
\label{sect.discussion}

We have examined the behaviour of the pseudo-marginal random walk
Metropolis algorithm in the limit as the dimension of the target
\mbox{approaches} infinity, under the assumption that the noise in the
estimate of the log-target at a proposed new value, $\mathbf{x}$, is additive
and independent of $\mathbf{x}$.

Subject to relatively general conditions
on the target, limiting forms for the acceptance rate and for the
efficiency, in terms of expected
squared jump distance (ESJD), have been obtained. We examined two
different noise distributions (Gaussian and Laplace),
and found that the optimal scaling of the proposal is
insensitive to the variance of the noise and to whether the noise has
a Gaussian or a Laplace distribution.

We then examined the behaviour of the Markov chain on the target,
$\mathbf{x}$, and the noise, obtaining a limiting diffusion for the first
component of a target with independent and identically distributed
components.
%The proof was very intricate due to
%the non-Markovian nature of the transformed process $V^{(d)}$.)
The efficiency function in this case is proportional to the speed of
the diffusion, thus
further justifying the use of ESJD in this context.

We identified a ``standard asymptotic regime'' under which
the additive noise is Gaussian with variance inversely proportional
to the number of unbiased estimates that are used. In this regime
the efficiency
function is especially tractable, and we showed that it is maximised when
the acceptance rate is approximately 7.0\% and the variance of the
Gaussian noise
is approximately 3.3. We noted that in this regime the optimal
noise variance is also insensitive to the choice of scaling.

A detailed simulation study on a Lotka--Volterra Markov jump process
using a particle filter suggested that in the scenario considered the
assumptions of the standard asymptotic regime are reasonable provided the
number of particles is not too low. Furthermore, whilst the assumption
that the
distribution of the noise does not depend on the current position is
not true, variations in the distribution have a small effect on the
distribution of the estimates of the log-target compared with the
effect of the
noise itself. The optimal scaling was found to be insensitive to the
noise variance (or equivalently the number of particles), and the
optimal noise variance was relatively insensitive to the choice of
scaling. The overall optimal scaling was consistent with the
theoretical value obtained; however the optimal variance was a little
lower than the theoretically optimal value. Investigations
showed that this discrepancy can be explained by the differences
between our
theoretical measure of efficiency (ESJD) and
empirical measures used in the simulation study (ESS).

The results from the simulation study suggest that in low dimension a
safer option than tuning to a
particular variance and acceptance rate might be to take advantage of the
insensitivity of the optimal scaling to the variance and vice versa
and optimise scaling and variance independently.

The diffusion limit provides strong support for the optimisation
strategies suggested by the ESJD criterion. However, in an ideal
world it would be good to show that the sequence of algorithms which
achieves the minimal optimal integrated autocorrelation time for a
given functional might converge to the optimal diffusion. This is
%of course
a generic question which is relevant to all diffusion limits for MCMC
algorithms, and there are still important open questions regarding the
relationships between ESJD, diffusion limits, and limiting optimal
integrated autocorrelation. In this direction, a recent paper \cite{RRcomplexity} has shown that diffusion limits can be translated into
\textit{complexity} results, thus demonstrating that at least the order of
magnitude of the number of iterations to ``converge'' can be read off
from the diffusion limit.

% We examined two noise scenarios in the case where there is no such
% product, and in both cases found that using only one unbiased estimate
% is, in fact, optimal. \cscomment{Gareth: explanation? Relate to work
% with Kasper?}.

The optimal variance of 3.28 under the standard asymptotic regime is
similar to the value of 2.83 obtained in
\cite{Doucetetal2014} under the same noise assumptions and for a
scenario where the component of the Markov chain on $\mathcal{X}$
mixes infinitely more
slowly than the noise component. Indeed, as noted in a remark
following Corollary~\ref{cor.max.Gauss}, 2.83 is (to two
decimal places)
the optimal variance that we obtain when $\ell=0$. There are many
differences between the approaches in \cite{Doucetetal2014} and this
article. For example, we optimise a limiting
efficiency for the random walk Metropolis with respect to both the
scaling and the variance whereas Doucet et al. \cite{Doucetetal2014}
consider the
univariate optimisation of a bound on the efficiency of
Metropolis--Hastings kernels which satisfy a positivity
condition. That a similar conclusion may be drawn from two very
different approaches is encouraging.

%sA #&#
\begin{appendix}
%sA #&#
\section{Proof of technical lemmas}

Let\vspace*{-1pt} $\{X_j\}_{j \geq1}$ be an i.i.d. sequence of random variables
distributed as $f(x)   \,dx$,
%$W \dist\Normal(\tau^2/2, \tau^2)$,
$W \stackrel{\mathcal{D}}{\sim}e^w g^*(w)   \,dw$,
$\{Z_{k,j}\}_{k \geq0, j \geq1}$
an i.i.d. sequence of $\mathbf{N}(0,1)$ random variables, $\{U_k\}_{k
\geq0}$ an i.i.d. sequence of random variables
uniformly distributed on $(0,1)$,
%and $\{W^{*}_k\}_{k \geq0}$ an i.i.d sequence of $\Normal(-\tau^2/2,
and $\{W^{*}_k\}_{k \geq0}$ an i.i.d. sequence distributed as $g^*(w)
  \,dw$.
All these random variables are assumed to be independent from one another.
For all integers $1 \leq j \leq d$ we set $X^{(d)}_{0,j}=X_j$ and
$W^{(d)}_0=W$. We introduce the proposals
$X^{(d),*}_{k,j}=X^{(d)}_{k,j} + \ell  I^{-1/2}   d^{-{1}/2}
Z_{k,j}$ and define $ (X^{(d)}_{k+1}, W^{(d)}_{k+1} )=
(X^{(d),*}_{k}, W^{*}_k )$ if
\[
U_k < F \Biggl( W^{*}_k-W^{(d)}_{k}
+ \sum_{j=1}^d A\bigl(X^{(d),*}_{k,j}
\bigr)-A\bigl(X^{(d)}_{k,j}\bigr) \Biggr)
\]
and $ (X^{(d)}_{k+1}, W^{(d)}_{k+1} )= (X^{(d)}_{k},
W^{(d)}_{k} )$ otherwise. We define $\mathbf{X}^{(d)} =
(X^{(d)}_{k,1}, \ldots, X^{(d)}_{k,d})$.
For any dimension $d \geq1$ the process $\{\mathbf{X}^{(d)}_{k},
W^{(d)}_{k})\}_{k \geq1}$ is\vspace*{-1pt} a Metropolis--Hastings
Markov chain started at stationarity, that is, $(\mathbf{X}^{(d)}_0,
W^{(d)}_0)=(X_1,\ldots,\break X_d,W) \stackrel{\mathcal{D}}{\sim}\pi
^{(d)}$, targeting the distribution $\pi^{(d)}$.

%sA.1 #&#
\subsection{Proof of Lemma \texorpdfstring{\protect\ref{lem.asymp.gen}}{1}}
\label{sec.asymp.gen.lem}
In this section, for notational convenience, we write $Z_j$ instead of
$Z_{0,j}$ and $W^*$ instead of $W^{*}_0$. We set
%
% \begin{array}{ll}
%a(w) &:= \EE\big[ F'(\Omega+ W^* - w) \big]\\
%b(w) &:= \frac{1}{2}   \EE\big[ F(\Omega+ W^* - w) \big]
% \end{array}
%
%eA.1 #&#
\begin{equation}
\label{eq.def.ab}
\hspace*{6pt}\quad a(w) := \mathbb{E} \bigl[ F'\bigl(\Omega+ W^* - w
\bigr) \bigr] \quad\mbox{and}\quad b(w) := \tfrac{1}{2} \mathbb{E} \bigl[ F
\bigl(\Omega+ W^* - w\bigr) \bigr]
\end{equation}
with $F'(u) = e^u   \mathrm{I}_{\{u<0\}}$ and $\Omega\stackrel{\mathcal{D}}{\sim}\mathbf{N}(-\ell^2/2, \ell^2)$ independent from
all other sources of randomness.
To prove Lemma~\ref{lem.asymp.gen}, it suffices to show that the
function $a$ and $b$
are continuous, bounded, satisfy identity \eqref{a.averaging.identity}, and that the following two limits hold:
%
%eA.2 #&#
\begin{equation}
\label{e.reduction}
\cases{
\ds\lim
_{d \to\infty} \mathbb{E} \bigl| \mathbb{E}_d \bigl[
\bigl(X^{1,d}_1-X_1\bigr) / \delta \bigr] -
\ell^2 I^{-1} a(W)A'(X_1)
\bigr|^2 = 0,
\vspace*{3pt}\cr
\ds\lim_{d \to\infty} \mathbb{E} \bigl| \tfrac{1}2 \mathbb
{E}_d \bigl[ \bigl(X^{1,d}_1-X_1
\bigr)^2 / \delta \bigr] - \ell^2 I^{-1} b(W)
\bigr|^2 = 0.}
\end{equation}
We have used the notation $\mathbb{E}_d[   \cdots]$ for $\mathbb
{E}[  \cdots  | X_1, \ldots,X_d,W]$.
The fact that the functions $a$ and $b$ are bounded and continuous
follows from the dominated convergence theorem.
\begin{itemize}
\item \textit{Proof of equation} \eqref{a.averaging.identity}.
% \red{[I suspect that there is a nicer proof]}
Note that $\mathbb{E}[b(W)] = \frac{1}2 \mathbb{E}[1 \wedge\exp
(\Omega+ B)]$ with $B := W^*-W$. A standard computation show that for
any $\beta\in\mathbf{R}$, we have $\mathbb{E}[1 \wedge\exp(\Omega
+ \beta)] = 2 \Phi(-\ell/2 + \beta/ \ell)$, so that the identity
$\mathbb{E}[b(W)] = \frac{1}{2} \alpha(\ell)$ directly follows from
the definition of $\alpha$ in Theorem~\ref{thm.asymp.analysis}.

For proving the identity $\mathbb{E}[a(W)] = \frac{1}2 \alpha(\ell)$,
note that the expectation $\mathbb{E}[a(W)]$ equals
\begin{eqnarray*}
&& \int \!\!\int \!\!\int_{(z,w,w^*) \in\mathbf{R}^3} e^{-\ell^2/2 + \ell z + w^*
- w}
\mathrm{I}_{\{-\ell^2/2 + \ell z + w^* - w < 0\}} e^{w} g^*(w)g^*\bigl(w^*\bigr)
\\
&&\hspace*{48pt}\qquad{}\times\frac{e^{-z^2/2}}{\sqrt{2 \pi}} \,dw \,dw^* \,dz
\\
&& \qquad = \int\!\!\int\!\!\int_{(z,w,w^*) \in\mathbf{R}^3} \mathrm{I}_{\{-\ell^2/2 + \ell
(-z+\ell) + w - w^* > 0\}}
e^{w^*} g^*(w)g^*\bigl(w^*\bigr)\\
&&\hspace*{80pt}\qquad{}\times \frac{e^{-(-z+\ell)^2/2}}{\sqrt{2 \pi}} \,dw \,dw^* \,dz
\\
&&\qquad = \int \!\!\int \!\! \int_{(z,w,w^*) \in\mathbf{R}^3}
\mathrm{I}_{\{-\ell^2/2 + \ell
z + w^* - w > 0\}} e^{w^*} g^*(w)g^*\bigl(w^*\bigr)
\\
&&\hspace*{80pt}\qquad{}\times\frac{e^{-z^2/2}}{\sqrt{2 \pi}} \,dw \,dw^* \,dz
\\
&&\qquad= \mathbb{E}[ \mathrm{I}_{\{\Omega+W^*-W > 0\}}].
\end{eqnarray*}
We have used the change of variable $(z,w^*,w) \to(-z + \ell, w,
w^*)$ to go from the second line to the third. This computation shows
that $\mathbb{E}[a(W)] := \mathbb{E}[ e^{\Omega+ W^*-W}   \mathrm
{I}_{\{\Omega+W^*-W < 0\}}]= \mathbb{E}[ \mathrm{I}_{\{\Omega+W^*-W
> 0\}}]$. Since $F(u) = 1 \wedge e^u = \break e^u   \mathrm{I}_{\{u<0\}} +
\mathrm{I}_{u \geq0}$, it follows that
\[
\alpha(\ell) = \mathbb{E}\bigl[F\bigl(\Omega+W^*-W\bigr)\bigr] = \mathbb{E}
\bigl[ e^{\Omega+ W^*-W} \mathrm{I}_{\{\Omega+W^*-W < 0\}}\bigr] + \mathbb{E}[
\mathrm{I}_{\{\Omega+W^*-W > 0\}}],
\]
and therefore $\mathbb{E}[a(W)] = \alpha(\ell)/2$.
\item \textit{Proof of equation} \eqref{e.reduction}.
We will only verify that the first limit in equation \eqref
{e.reduction} holds. The proof of the second limit is similar but
easier. In other words,
we will focus on proving that the sequence $\mathbb{E}_d
[(X^{1,d}_1-X_1) / \delta ]$ converges in $L^2$ to $\ell^2
I^{-1}   a(W)A'(X_1)$. An integration by parts shows that for any
continuous function $g\dvtx \mathbf{R}\to\mathbf{R}$ such that $g'$ has a
finite number of discontinuities, if $g(Z)$ and $g'(Z)$ have a finite
first moment for $Z \stackrel{\mathcal{D}}{\sim}\mathbf{N}(0,1)$,
the identity $\mathbb{E}[Z \times g(Z)] = \mathbb{E}[g'(Z)]$ holds.
It follows that
\begin{eqnarray*}
&& \mathbb{E}_d \bigl[ \bigl(X^{1,d}_1 -
X_1\bigr)/\delta \bigr]\\
&&\qquad = \ell I^{-1/2}
\delta^{1/2} \mathbb{E}_d\bigl[ Z_1 \times F
\bigl(\Omega^{(d)} + W^* - W\bigr) \bigr]
\\
&&\qquad= \ell^2 I^{-1} \mathbb{E}_d \bigl[
F'\bigl(\Omega^{(d)} + W^* - W\bigr) \times
A'\bigl(x_1 + \ell I^{-1/2}
\delta^{1/2} Z_1\bigr) \bigr]
\end{eqnarray*}
with $\Omega^{(d)} = \sum_{i=1}^d A(X_i + \ell  I^{-1/2}   \delta
^{1/2}   Z_i) - A(X_i)$. Under Assumption~\ref{assump.f} the function
$A' = (\log f)'$ is globally Lipschitz so that, since the function $F'$
is bounded, one can focus on proving that
\[
\mathbb{E}_d \bigl[ F'\bigl(\Omega^{(d)} +
W^* - W\bigr) \bigr] \times A'(X_1)
\]
converges in $L^2$ to $a(W)A'(X_1)$. By the Cauchy--Schwarz inequality,
this reduces to proving that
\[
\lim_{d \to\infty} \mathbb{E} \bigl[ \bigl|\mathbb{E}_d\bigl[
F'\bigl(\Omega^{(d)} + W^* - W\bigr) \bigr] -
\mathbb{E}_d\bigl[ F'\bigl(\Omega+ W^* - W\bigr) \bigr]
\bigr|^4 \bigr] = 0.
\]
By the Portmanteau's theorem, the dominated convergence theorem, and
the definition of $\Omega^{(d)}$, this reduces to proving that for
almost every realisation $\{x_i\}_{i \geq1}$ of the i.i.d. sequence $\{
X_i\}_{i \geq1}$ the following limit holds in distribution:
\[
\lim_{d \to\infty} \sum_{i=1}^d
A\bigl(x_i + \ell I^{-1/2} \delta^{1/2}
Z_i\bigr) - A(x_i) = \Omega.
\]
Under Assumption~\ref{assump.f} the third derivative of $A$ is bounded
so that a second order Taylor expansion yields that the difference
$A(x_i + \ell  I^{-1/2}   \delta^{1/2}   Z_i) - A(x_i) $ equals
$A'(x_i) \ell  I^{-1/2}   \delta^{1/2}   Z_i + (1/2)
A^{\prime\prime}(x_i)  \ell^2   I^{-1}   \delta  Z^2_i + \mathcal
{O}(d^{-3/2})$; consequently,
\begin{eqnarray*}
&& \sum_{i=1}^d A\bigl(x_i +
\ell I^{-1/2} \delta^{1/2} Z_i\bigr) -
A(x_i) \\
&&\qquad \stackrel{\mathrm{law}}{=} \frac{\ell^2}{2   I}
\biggl\{ \frac{\sum_{i=1}^d
A^{\prime\prime}(x_i)}{d} \biggr\} +\ell I^{-1/2} \biggl\{
\frac{\sum_{i=1}^d A^{\prime}(x_i)^2}{d} \biggr\}^{1/2} \xi
\\
&&\hspace*{3pt}\qquad\quad{}+ \frac{\ell^2}{2   I} \biggl\{ \frac{\sum_{i=1}^d
A^{\prime\prime}(x_i)   (Z_i^2-1)}{d} \biggr\} + \mathcal{O}
\bigl(d^{-1/2}\bigr)
\end{eqnarray*}
for $\xi\stackrel{\mathcal{D}}{\sim}\mathbf{N}(0,1)$ independent
from all other sources of randomness. The law of large numbers shows
that for almost every realisation $\{x_i\}_{i \geq1}$ the right-hand
side of the above equation converges in distribution towards $\Omega
\stackrel{\mathcal{D}}{\sim}\mathbf{N}(-\ell^2/2, \ell^2)$.
\end{itemize}

%sA.2 #&#
\subsection{Proof of Lemma \texorpdfstring{\protect\ref{lem.separation.scale}}{2}}
\label{sec.sep.sc.lem}
For convenience, we first give a high-level description of the reasoning.
We construct processes $\{\widehat{W}^{(d)}_{k}\}_{k \geq0}$, $\{
\widehat{Y}^{(d)}_{k}\}_{k \geq0}$, and $\{Y_{k}\}_{k \geq}$
satisfying the following:
\begin{itemize}
\item With high probability $\widehat{W}^{(d)}_{k} = W^{(d)}_{k}$ for
all $k \leq T_d$.
\item The process $\{\widehat{Y}^{(d)}_{k}\}_{k \geq0}$ has the same
law as the process $\{\widehat{W}^{(d)}_{k}\}_{k \geq0}$.
\item With high probability $\widehat{Y}^{(d)}_{k} = Y_{k}$ for all $k
\leq T_d$.
\item The process $\{Y_{k}\}_{k \geq0}$ is a Markov chain that is
ergodic with invariant distribution $e^w   g^*(w)   \,dw$.
\end{itemize}
One can thus use an approximation of the type
\[
\mathbb{E} \Biggl| \frac{1}{T_d} \sum_{k=0}^{T_d-1}
h\bigl(W^{(d)}_{k}\bigr) - \mathbb{E}\bigl[h(W)\bigr] \Biggr|
\approx \mathbb{E} \Biggl| \frac{1}{T_d} \sum_{k=0}^{T_d-1}
h(Y_{k}) - \mathbb {E}\bigl[h(W)\bigr] \Biggr|
\]
and the usual ergodic theorem gives the conclusion.
We use at several places the following elementary lemma.

%le4 #&#
\begin{lemma} \label{lem.coupling.traj}
Let $T_d = \lfloor{d^{\gamma} } \rfloor$ with $0 < \gamma<\frac{1}4$.
Let $\{ P^{(d)}_{k} \}_{k,d \geq0}$ and $\{Q^{(d)}_{k}\}_{k,d \geq0}$
be two arrays of $(0,1)$-valued random variables.
Let $\{U_k\}_{k \geq0}$ be a sequence of random variables uniformly
distributed on the interval $(0,1)$. We suppose that for all dimension
$d \geq1$ the random variable $U_k$ is independent from $\{P^{(d)}_j\}
_{j=0}^{k-1}$ and $\{Q^{(d)}{j}\}_{j=0}^{k-1}$. Consider the event
\[
E^{(d)}_k := \bigl\{ \omega\dvtx  \mathrm{I}_{ \{U_j < P^{(d)}_j \} } =
\mathrm{I}_{ \{U_j < Q^{(d)}_j
\} } \mbox{ for all } 0 \leq j \leq k \bigr\}.
\]
Under the assumption that
$\mathbb{E} [ |P^{(d)}_{k} - Q^{(d)}_{k}|    |
E^{(d)}_{k-1} ] \lesssim k / \sqrt{d}$,
we have
\[
\lim_{d \to\infty}   \mathbb{P} \bigl( E^{(d)}_{T_d}  \bigr) = 1.
\]
\end{lemma}

\begin{pf}
Note that $\mathbb{P}(E^{(d)}_{k}) = \mathbb{P}(E^{(d)}_{0})   \prod_{j=1}^k \mathbb{P} [ \mathrm{I}_{ \{U_j < P^{(d)}_{j} \} } =
\mathrm{I}_{ \{U_j < Q^{(d)}_j \} }    | E^{(d)}_{j-1} ]$.
Since $U_j$ is supposed to be\vspace*{1pt} independent from the event
$E^{(d)}_{j-1}$, it follows that
$\mathbb{P} [ \mathrm{I}_{ \{U_j < P^{(d)}_{j} \} } = \mathrm
{I}_{ \{U_j < Q^{(d)}_j \} }    | E^{(d)}_{j-1} ] = 1- \mathbb
{E} [ |P^{(d)}_{j} - Q^{(d)}_{j}|    | E^{(d)}_{j-1} ]$.
The conclusion then directly follows from the bound
$\mathbb{E} [ |P^{(d)}_{k} - Q^{(d)}_{k}|    |
E^{(d)}_{k-1} ] \lesssim k / \sqrt{d}$ and $\gamma< 1/4$.
\end{pf}
We now describe the construction of the processes $\{\widehat
{W}^{(d)}_{k}\}_{k \geq0}$, $\{\widehat{Y}^{(d)}_{k}\}_{k \geq0}$
and $\{Y_{k}\}_{k \geq0}$. To this end, we need
an i.i.d. sequence $\{\xi_k\}_{k \geq0}$ of standard $\mathbf{N}(0,1)$ Gaussian random variables independent from all other sources
of randomness.
All the processes start at the same position $W^{(d)}_0 = \widehat
{W}^{(d)}_0= \widehat{Y}^{(d)}_0=Y_0=W$.
We define $\widehat{W}^{(d)}_{k+1}=W^{*}_k$ if
\[
U_k < F \Biggl[ \frac{\ell}{\sqrt{d   I}}\sum
_{j=1}^d A'(X_j)
Z_{k,j} - \ell^2/2 + W^{*}_k -
\widehat{W}^{(d)}_{k} \Biggr]
\]
and $\widehat{W}^{(d)}_{k+1}=\widehat{W}^{(d)}_{k}$ otherwise. We
define $\widehat{Y}^{(d)}_{k+1}=W^{*}_k$ if
\[
U_k < F \Biggl[ \ell I^{-1/2} \Biggl\{ d^{-1}\sum
_{j=1}^d A'(X_j)^2
\Biggr\}^{1/2} \xi_k - \ell^2/2 +
W^{*}_k - \widehat{Y}^{(d)}_{k}
\Biggr]
\]
and $\widehat{Y}^{(d)}_{k+1}=\widehat{Y}^{(d)}_{k}$ otherwise. We
define $Y_{k+1}=W^{*}_k$ if
\[
U_k < F \bigl[ \ell \xi_k - \ell^2/2 +
W^{*}_k - Y_{k} \bigr]
\]
and $Y_{k+1}=Y_{k}$ otherwise.
\begin{itemize}
\item $W^{(d)}_{k} = \widehat{W}^{(d)}_{k}$ \textit{with high
probability}.
We prove that $\lim_{d \to\infty}   \mathbb{P} [ W^{(d)}_{k} =\break 
\widehat{W}^{(d)}_{k}   \dvtx    k=1, \ldots, T_d  ] = 1$.
Because the Metropolis--Hastings function $F$ is globally Lipschitz,
Lemma~\ref{lem.coupling.traj} shows that it suffices to verify
that
%
%eA.3 #&#
\begin{equation}
\label{e.boun.taylor}
\hspace*{6pt}\mathbb{E} \Biggl| \sum_{j=1}^d
A\bigl(X^{(d),*}_{k,j}\bigr)-A\bigl(X^{(d)}_{k,j}
\bigr)-A'(X_j) \ell I^{-1/2} Z_{k,j}
/ \sqrt{d} + \frac{\ell^2}{2} \Biggr| \lesssim k / \sqrt{d}.
\end{equation}
Under Assumption \ref{assump.f} the second and third derivatives of
$A$ are bounded so that bound \eqref{e.boun.taylor}
follows from a second-order Taylor expansion,
\begin{eqnarray*}
&& \mathbb{E} \Biggl| \sum_{j=1}^d  A
\bigl(X^{(d),*}_{k,j}\bigr)-A\bigl(X^{(d)}_{k,j}
\bigr)-A'(X_j) \ell I^{-1/2} Z_{k,j}
/ \sqrt{d} + \ell^2 / 2 \Biggr|
\\
&&\!\qquad\lesssim \mathbb{E} \Biggl| \sum_{j=1}^d A
\bigl(X^{(d),*}_{k,j}\bigr)-A\bigl(X^{(d)}_{k,j}
\bigr)-\frac{\ell}{\sqrt{d
I}}A'\bigl(X^{(d)}_{k,j}
\bigr)Z_{k,j} - \frac{\ell^2}{2   I
d}A''
\bigl(X^{(d)}_{k,j}\bigr) Z_{k,j}^2 \Biggr|
\\
&&\!\qquad\quad{}+ \frac{\ell}{\sqrt{d   I}}\mathbb{E} \Biggl| \sum_{j=1}^d
\bigl(A'\bigl(X^{(d)}_{k,j}\bigr)-A'(X_j)
\bigr) Z_{k,j} \Biggr| \\
&&\!\qquad\quad{}+ \frac{\ell^2}{2   I   d}\mathbb{E} \Biggl| \sum
_{j=1}^d \bigl(A''
\bigl(X^{(d)}_{k,j}\bigr)-A''(X_j)
\bigr) Z_{k,j}^2 \Biggr|
%&&\qquad\quad{}
+ \frac{\ell^2}{2   I} \mathbb{E} \Biggl| \frac{1}{d}\sum
_{j=1}^{d} A''(X_j)
+ I \Biggr|
\\
&&\!\qquad\lesssim \frac{1}{\sqrt{d}} + \frac{1}{\sqrt{d}} \Biggl\{ \sum
_{j=1}^d\mathbb{E} \bigl|A'
\bigl(X^{(d)}_{k,j}\bigr)-A'(X_j)
\bigr|^2 \Biggr\}^{1/2} \\
&&\!\qquad\quad{}+ \frac{1}{2d} \sum
_{j=1}^d \mathbb{E} \bigl| A''
\bigl(X^{(d)}_{k,j}\bigr)-A''(X_j)
\bigr|
%&&\qquad\quad{}
+ \mathbb{E} \Biggl| d^{-1}\sum_{j=1}^{d}
A''(X_j) + I \Biggr| \\
&&\!\qquad\lesssim
\frac{1}{\sqrt{d}} + \frac{k}{\sqrt{d}} + \frac{k}{\sqrt
{d}} + \frac{1}{\sqrt{d}}.
\end{eqnarray*}
We have used the bound $\mathbb{E}|X^{(d)}_{k,j}-X_j|^2 \lesssim\frac
{k^2}{d}$.

\item \textit{$\widehat{W}^{(d)}$ and $\widehat{Y}^{(d)}$ have same
law.}
It is straightforward to verify that the processes $\{\widehat
{W}^{(d)}_{k}\}_{k \geq0}$ and $\{\widehat{Y}^{(d)}_{k}\}_{k \geq0}$
have the same law.

\item \textit{$ \widehat{Y}^{(d)}_{k} = Y_{k}$ with high probability.}
We prove that $\lim_{d \to\infty}   \mathbb{P} [ \widehat
{Y}^{(d)}_{k} = Y_{k}   \dvtx    k=1, \ldots, T_d  ] = 1$.
Lemma~\ref{lem.coupling.traj} shows that this follows from the
elementary bound $\mathbb{E} |  \{ d^{-1}\sum_{j=1}^d
A'(X_j)^2  \}^{1/2} - I^{1/2}  | \lesssim1 / \sqrt{d}$.
%One possible proof consists in exploiting the fact
%that the function $x \mapsto\sqrt{I+x}$ is Lipschitz
%on the domain $(-I/2;\infty)$, that
%$\EE\big| d^{-1}\sum_{j=1}^d A'(X_j)^2 -
%I \big| \lesssim1 / \sqrt{d}$ and that (Markov's inequality)
%the quantity $\Big\{ d^{-1}\sum_{j=1}^d A'(X_j)^2 - I \Big\}$
% does not belong to $(-I/2;\infty)$ with probability at
%most $\frac{4   \mathrm{Var}\big(A'(X)^2 \big)}{d \times I^2}$.
\end{itemize}

We now show that the Markov chain $\{Y_{k}\}_{k \geq0}$ is a Markov
chain that is reversible with respect to the distribution $e^w g^*(w)\,dw$,
\[
e^x g^*(x) g^*(y) \mathbb{E} \bigl[ \mathbb{E}\bigl[F(\Omega+ y -
x)\bigr] \bigr] = e^y g^*(y) g^*(x) \mathbb{E} \bigl[ \mathbb{E}
\bigl[F(\Omega + x - y)\bigr] \bigr]
\]
for all $x,y \in\mathbf{R}^2$.
This boils down to verifying that the function $(x,y) \mapsto e^x
\mathbb{E}[F(\Omega+ y - x)]$ is symmetric; Proposition $2.4$ of
\cite{RobertsGelmanGilks1997} shows that this quantity can be
expressed as
\[
e^x \Phi \biggl( \frac{-({1}/2) \ell^2 + y - x}{\ell} \biggr)+e^y \Phi
\biggl( \frac{-({1}/2) \ell^2 + x - y}{\ell} \biggr),
\]
which is indeed symmetric. Note that this Markov chain corresponds to
the \textit{penalty method} of \cite{ceperley1999penalty}; see also \cite
{nicholls2012coupled} for a discussion of this algorithm.
The ergodic theorem for Markov chains applies; for any bounded and
measurable function $h\dvtx \mathbf{R}\to\mathbf{R}$ we have
\[
\lim_{N \to\infty} \mathbb{E} \Biggl| \frac{1}{N}\sum
_{k=0}^{N-1} h(Y_{k}) - \mathbb{E}
\bigl[h(W)\bigr] \Biggr|=0.
\]
One can thus use the triangle inequality several times to see that for
any bounded and measurable function $h\dvtx \mathbf{R}\to\mathbf{R}$, we have
\begin{eqnarray*}
&&\mathbb{E} \Biggl| \frac{1}{T_d} \sum_{k=0}^{T_d-1}
 h\bigl(W^{(d)}_{k}\bigr) - \mathbb{E}\bigl[h(W)\bigr] \Biggr|
\\
&&\qquad\leq   \frac{1}{T_d} \sum_{k=0}^{T_d-1}
\mathbb{E} \bigl| h\bigl(W^{(d)}_{k}\bigr) - h\bigl(
\widehat{W}^{(d)}_{k}\bigr) \bigr| + \mathbb{E} \Biggl|
\frac{1}{T_d} \sum_{k=0}^{T_d-1} h\bigl(
\widehat {W}^{(d)}_{k}\bigr) - \mathbb{E}\bigl[h(W)\bigr] \Biggr|
\\
&&\qquad\lesssim  \bigl(1-\mathbb{P}\bigl[W^{(d)}_{k} =
\widehat{W}^{(d)}_{k} \dvtx  k=1, \ldots, T_d\bigr]
\bigr) + \mathbb{E} \Biggl| \frac{1}{T_d} \sum_{k=0}^{T_d-1}
h\bigl(\widehat{Y}^{(d)}_{k}\bigr) - \mathbb{E}\bigl[h(W)
\bigr] \Biggr|
\\
&&\qquad\lesssim  o(1) + \frac{1}{T_d} \sum_{k=0}^{T_d-1}
\mathbb{E} \bigl| h\bigl(\widehat{Y}^{(d)}_{k}\bigr) -
h(Y_{k}) \bigr| + \mathbb{E} \Biggl| \frac{1}{T_d} \sum
_{k=0}^{T_d-1} h(Y_{k}) - \mathbb{E}
\bigl[h(W)\bigr] \Biggr|
\\
&&\qquad= o(1) + o(1) + o(1),
\end{eqnarray*}
which completes the proof of Lemma~\ref{lem.separation.scale}.

%sA #&#
\section{Details of the Lotka Volterra model}
\label{sec.lotka.details}

In this Appendix, we present details of the Lotka--Volterra model
used in the simulation study of Section~\ref{sect.sim.study}. The
Lotka--Volterra model is a continuous-time Markov chain on $\mathbb{N}_0^2$.
The transitions and associated rates for this model are
\begin{eqnarray*}
(u_1,u_2) & \stackrel{x_1u_1u_2}{\longrightarrow}  & (u_1+1,u_2-1), \qquad (u_1,u_2)
\stackrel{x_2u_1} {\longrightarrow} (u_1-1,u_2)
\quad\mbox{and}\\
 (u_1,u_2) & \stackrel{x_3u_2}
{\longrightarrow} & (u_1,u_2+1);
\end{eqnarray*}
the rate for any other transition is zero. Observations of the Markov
chain, when they occur, are subject to
Gaussian error,
\[
\mathbf{Y}(t)\sim\mathbf{N}\lleft( \lleft[ %
\begin{array}
{c} u_1(t)
\\
u_2(t) \end{array}
 \rright], \lleft[
\matrix{ x_4&0
\cr
0&x_5 }
 \rright] \rright).
\]
Using $\mathbf{x}=(0.006,0.6,0.3,25,49)$,
a realisation of the stochastic process was simulated from initial value
$\mathbf{u}(0)=(70,70)$ for $T=50$ time units. The state, perturbed with
Gaussian noise, $\mathbf{y}(t)$, was recorded at
$t=1,2,\dots,T$.
For inference, $X_1,\ldots,X_5$ were assumed to be
independent, \textit{a priori} with $\log
X_i \sim \operatorname{Unif}[-8,8]$, ($i=1,\ldots,5$).

The initial value for each chain was a vector of estimates of the posterior
median for each parameter, obtained from the initial run; hence no
``burn-in'' was required.
Each algorithm was run for $2.5\times10^5$
iterations, except with $m=50$ and $m=80$, where $10^6$ iterations were
used. Output was thinned by a factor of $10$ for storage.
\end{appendix}

\section*{Acknowledgements}
We are grateful to the Associate Editor and three referees for their
comments, which helped improve both the presentation and the content
of this article.
Gareth Roberts and Jeffrey Rosenthal are grateful for financial support in carrying out this research from,
respectively, EPSRC of the UK, through the CRiSM (EP/D002060/1) and iLike (EP/K014463/1)
projects, and NSERC of Canada.

%

% imsref loaded by daiva.urboniene, 2014-10-27 14:48:12
% imsref loaded by daiva.urboniene, 2014-10-27 15:11:34


%suskaldyti doi

\printaddresses
\end{document}